\documentclass[11pt]{article}
\usepackage{graphicx}

\def\be{\begin{equation}}
\def\ee{\end{equation}}
\def\ba{\begin{eqnarray}}
\def\ea{\end{eqnarray}}

\def\12{{1\over 2}}

\def\msun{M_\odot}

\def\etal{{\it et~al.~}}
\def\ltsima{$\; \buildrel < \over \sim \;$}
\def\simlt{\lower.5ex\hbox{\ltsima}}
\def\gtsima{$\; \buildrel > \over \sim \;$}
\def\simgt{\lower.5ex\hbox{\gtsima}}

\def\zsun{Z_\odot}

\begin{document}
% \selectlanguage{english}

\title{\bf Dynamics of a Supernova Envelope in a Cloudy Interstellar Medium\footnote{This paper is published
in Astronomy Reports, 2015, Vol. 59, No. 7, pp. 690.}}
\author{V.~V.~Korolev$^1$, E.~O.~Vasiliev$^{2}$\thanks{eugstar@mail.ru}, I. G. Kovalenko$^{1}$, Yu.~A.~Shchekinov$^{3}$ \\
\it $^1$Volgograd State University, Volgograd, Russia \\
\it $^2$Institute of Physics, Southern Federal University, Rostov-on-Don, Russia \\
\it $^3$Physics Department, Southern Federal University, Rostov-on-Don, Russia}

\date{}

\maketitle

\begin{abstract}
The evolution of a supernova remnant in a cloudy medium as a function of the volume filling
factor of the clouds is studied in a three-dimensional axially symmetrical model. The model includes the
mixing of heavy elements (metals) ejected by the supernova and their contribution to radiative losses.
The interaction of the supernova envelope with the cloudy phase of the interstellar medium leads to nonsimultaneous,
and on average earlier, onsets of the radiative phase in different parts of the supernova
envelope. Growth in the volume filling factor $f$ leads to a decrease in the time for the transition of the
envelope to the radiative phase and a decrease in the envelope’s mean radius, due to the increased energy
losses by the envelope in the cloudy medium. When the development of hydrodynamical instabilities in the
supernova envelope is efficient, the thermal energy falls as $E_t\sim t^{-2.3}$, for the propagation of the supernova
remnant through either a homogeneous or a cloudy medium. When the volume filling factor is $f\simgt 0.1$, a
layer with excess kinetic energy andmomentumforms far behind the global shock front from the supernova,
which traps the hot gas of the cavity in the central part of the supernova remnant. Metals ejected by the
supernova are also enclosed in the central region of the remnant, where the initial (high) metallicity is
essentially preserved. Thus, the interaction of the supernova envelope with the cloudy interstellar medium
appreciably changes the dynamics and structure of the distribution of the gas in the remnant. This affects
the observational characteristics of the remnant, in particularly, leading to substantial fluctuations of the
emission measure of the gas with $T>10^5$~K and the velocity dispersion of the ionized gas.
\end{abstract}

%\newpage

%----------------------- Section 1 -------------------------------

\section{Introduction}

\noindent

It is well known that the interstellar medium in galaxies is inhomogeneous and turbulent. An appreciable role in maintaining the turbulent flows is played by supernova explosions (see, e.g., \cite{scalo04}). The density contrast in these inhomogeneous media relative to the mean density ranges from $\sim$ 1 (in a diffuse medium) to 1000 or more (in molecular clouds). The interaction of shocks from supernovae with density inhomogeneities -- clouds -- can give rise to compression, rarefaction, vaporization, and acceleration of these inhomogeneities \cite{snclouds}. The effects of the reverse influence of the clouds on the dynamics of the supernova shocks are also evident. In spite of numerous numerical investigations of the disruption of individual clouds \cite{klein94} and ensembles of clouds \cite{polud02}, some questions and details of the interaction process have not been fully studied; in particular, the dependence of the remnant dynamics on the number of clouds in the interstellar medium -- the filling factor -- has remained unclear.

The overall pattern of the evolution of the remnant can depend appreciably on the characteristics of the inhomogeneities in themedium into which it expands. This became clear in the very first studies of the dynamics of flows with mass loading, specifically flows with matter added to the general flow of gas in the form of essentially stationary clouds \cite{dyson88}. In the case of shocks from supernovae, such tendencies can arise in a cloudy medium with density-enhancement factors in the clouds above 1000 in the near vicinity of the supernova, or about 10–100 far from the supernova, when the supernova shock becomes fairly weak, in order not to fully disrupt the clouds. The character
of the flows in intermediate regimes can vary during their evolution: some of the clouds that pass through the front earliest are disrupted, and the front itself weakened, while clouds located upstream in the flow are only weakly eroded by the shock, adding a
small fraction of their mass to the gaseous flow. The fragments of the disrupted clouds collide, partially merging and partially becoming more fully disrupted, leading to the appearance of turbulent flows far behind the global shock front, and a phase of mixing begins
in the gaseous flow \cite{polud02}. Layers with different dynamics, structure, thermal characteristics, and chemical composition can form. As a consequence, the shocks can vary the ionizational and molecular composition of the matter (see, e.g., \cite{hmk89,inutsuka-cloud00}), and the loaded flows themselves are subject to hydrodynamical instability \cite{s96}.

The influence of cooling of the gas is of special interest for the propagation of a shock front through a cloudy medium. As was noted above, shocks can destroy molecules and ionize gas, thereby influencing the thermal-loss rate and the subsequent chemical–dynamical evolution of the clouds and the entire gas flow. Individual regions of the initially adiabatic front become radiative at different times. Depending on the size and density of the clouds ahead of a front, the kinetic energy and momentum of the ordered radially expanding front from the supernova will be lost in interactions with these clouds. Moreover, after the perturbation of their outer layers by a shock front,
clouds that are in quasi-equilibrium with the external medium can either be compressed and become protostellar clouds, or expand, adding gas to the overall flow behind the front.

The process of enriching the interstellar medium in heavy elements -- metals -- is no less important. In supernovae, metals are ejected into the interstellar gas and redistributed in this gas on large scales. It is usually implicitly assumed that these metals mix fairly rapidly with the gas. However, both theoretical and observational studies over the past decade have shown that the mixing process can occupy a long time, depending on the mechanisms operating and the properties of the ambient medium. A low efficiency of the mixing is indicated by observations of inhomogeneities in the distributions of deuterium \cite{jenkins99}, oxygen \cite{meyer98,luck06}, and other elements \cite{bochkarev,satterfield} over broad spatial scales.

It is obvious that turbulent flows and cooling and heating of the gas play important roles in the redistribution and mixing of metals in the interstellar medium. Mixing in the interstellar medium was first studied numerically in \cite{avillez}: the characteristic time scale for mixing of chemical inhomogeneities in a medium excited by numerous fairly frequent supernovae proved to be of the order of 100 million years. The mixing of metals is also important for the intergalactic medium \cite{mix00,mix04,mix08,mix09}, where, in contrast to the interstellar medi\-um, there is no pumping energy from frequent supernovae. It is obvious that the mixing will be incomplete in regions with a low star-formation rate and rare supernovae. The efficiency of mixing in the case of a single supernova that explodes in a
medium without density variations is also low, and is limited by the early stages of the expansion, which have a fairly high Mach number \cite{mix12}. However, since the existence of appreciable inhomogeneities (clouds) in the medium inevitably changes the character of
the flow, these changes also affect the redistribution of heavy elements. We expect that turbulent flows arising during the disruption of the clouds will facilitate mixing of the cloud material with heavy elements ejected by supernovae.

Supernova remnants can be divided into two regions with different physical and dynamical properties: the hot cavity and the envelope. We will take the cavity to refer to hot gas in the central part of the remnant, which also contains gas highly enriched in heavy elements. The supernova envelope is formed of interstellar gas that is swept up by the shock from the supernova. Thus, we must bear in mind that
density inhomogeneities in the interstellar medium (clouds) could add material to both the envelope -- when they are torn apart by the global shock front from the supernova -- and the hot cavity -- if a cloud is not fully disrupted after it has passed through the
shock front.

The propagation of a supernova envelope in a two-phase interstellar medium is closely related to the formation of powerful galactic outflows or winds (see, e.g., \cite{wind-rev05,recchi-wind,sutherland-wind}). In this case, numerous coherent supernovae should form a common superbubble that moves through the inhomogeneous interstellar medium and breaks out of the galactic disk \cite{wind13}. It is obvious that clumpiness of the interstellar medium will increase the dissipation of energy and momentum, change the properties of the flow in the forming hot cavity and the surrounding superenvelope, and substantially lower the efficiency of their transformation into large-scale vertical outflows of interstellar gas \cite{s96,suchkov96,silich96}. Thus, studying the dynamics of a supernova envelope in a cloudy medium is also very important from the point of view of the large-scale dynamics of the interstellar medium. Calculations of the dynamics of a supernova remnant in a cloudy medium in a three-dimensional formulation were carried out in the recent study \cite{sn-turb14}, which was aimed at a detailed analysis of the radial distributions of momentum and energy as functions of the properties of the ambient interstellar medium. In our current study, we concentrated mainly on the dependences of the remnant dynamics and its observational properties on the contribution of clouds to the total density of the interstellar medium, more specifically, the volume
filling factor for the medium. This requires numerical experiments carried out in a wide interval of parameters for models of a cloudy medium corresponding to some filling factor. For this reason, we were limited to a three-dimensional axially symmetrical formulation.

We have studied the dynamics of a supernova remnant in a cloudy medium as a function of the volume filling factor of the clouds, taking into account the redistribution (mixing) of heavy elements ejected by the supernova into the ambient medium and their
influence on radiative losses. We will consider the dynamics of the mixing and the statistical properties of the distribution of heavy elements in a separate study, and restrict our analysis here to a description of the general characteristics required to compute the
radiative cooling.

%----------------------- Section 2 -------------------------------

\section{Model and numerical methods}

\noindent

\subsection{Gas Dynamics}

The following system of gas-dynamical equations is described in the model considered for the motion of the interstellar gas:
\begin{equation}
    {\partial \rho\over\partial t}+ \nabla(\rho \textbf{v})= 0,
\end{equation}
\begin{equation}
   {\partial \rho_{m}\over\partial t}+ \nabla(\rho_{m} \textbf{v})= 0,
\end{equation}
\begin{equation}
    {\partial (\rho \textbf{v}) \over\partial t}  + \nabla(\rho \textbf{v} \otimes \textbf{v} + p) = 0,
\end{equation}
\begin{equation}
    {\partial E\over\partial t} + \nabla\left((E+p)\textbf{v}\right)=
    n\Gamma - n^2\Lambda(T),
\end{equation}
\begin{equation}
     E ={\rho v^2\over 2} + {p\over \gamma-1}\;,
\end{equation}
\begin{equation}
     p={n k_B T },
\end{equation}
where $\rho$ and $\rho_m$ are the densities of the gas and metals; $n$ is the gas number density, $p$ is the pressure, $v$ is the velocity, $E$ is the total energy per unit volume of the gas, $T$ is the thermodynamic temperature, $\Lambda(T)$ is the cooling function, $\Gamma$ is the heating function, and $\gamma = 5/3$ is the adiabatic index. The metallicity of the gas in absolute units is given by the radio of the densities of the metals and gas $Z = \rho_m / \rho$. The dynamics of the metals are described using the transport of a passive scalar component that has the same velocity field as the gas \cite{avillez,mix04}.

We have studied this problem in an axially symmetric approximation, in which all parameters of the gas are specified in cylindrical coordinates and are assumed to be functions of the form $f=f(r, z)$. The solution of the system of equations presented above
was obtained numerically using an explicit, finite volume scheme without splitting the fluxes of quantitites in space using a TVD (Total Variation Diminishing) condition. This is a so-called MUSCL (Monotonic Upstream-Centered Scheme for Conservation Laws) scheme \cite{harten78,vanleer79,toro97,kulik02}; we used the Harten–Lax–van Leer–Contact (HLLC) approximation method to enhance the accuracy when computing fluxes at the cell boundaries \cite{vanleer77a,vanleer77b,vanleer79}. This approach has third-order accuracy in space in regions of smooth flows, and first-order accuracy at jumps, which facilitates stable and correct computation of trans-sonic regimes and flows with shocks. The method has second-order accuracy in time, which is achieved through the use of a Runge–Kutta stepwise computational
scheme. The source terms responsible for thermal processes were taken into account by separating them according to physical processes; i.e.,
the solution of the initial system of equations was obtained via a joint solution of the advection equations and the equation for the variation in the thermal energy due to radiative heating and cooling.

\subsection{Thermal Processes}

In ametal-enriched collisional gas with metallicity $Z\simgt 0.1~\zsun$, the contribution from cooling by metals becomes dominant when $T\simlt 10^7$~K (see, e.g., \cite{wiersma,v11}). Cooling of gas in the remnant is insignificant until the radiative cooling time scale becomes shorter than the age of the envelope, after which the gas begins to lose thermal energy. Before the onset of the radiative phase, the characteristic time scales for the ionization and recombination of metal ions in the remnant are shorter than the age of the envelope, so that the ionic composition of the gas is in collisional equilibrium. This collisional equilibrium is disrupted as the gas cools, since the recombination time for metal ions becomes longer than the cooling time. Thus, the gas remains in an ionization state corresponding
to a higher temperature than in the case of collisional equilibrium. This limits the approximation of ionization equilibrium, and a cooling function obtained for the time-dependent ion composition of the gas -- a non-equilibrium cooling function \cite{v11,v13} -— must be used after the onset of the radiative phase.

The computation of non-equilibrium gas cooling functions requires the solution of a system of more than 90 differential equations for each of the ion states of the main elements. Further, the sum of all the energy losses due to collisional ionization, excitation, recombination etc. must be found for the ion composition obtained. It is obvious that such a self-consistent computation of the ionization kinetics and the cooling functions for a two- or three-phase (metal-enriched) interstellar medium is currently difficult to realize in a multi-dimensional gas-dynamics approach, due to the huge amount of computational time this would require. Therefore, pre-calculated
tables for the cooling functions are used, in which the cooling rates are presented for only a few (from three to eight) gas metallicities (see, e.g., \cite{sd93}). This is clearly insufficient for fully correct investigations of the thermal evolution of the gas in media undergoing mixing, given the non-linear dependence of the cooling functions on the metallicity (see, e.g., \cite{sd93}).
Therefore, we used the method of \cite{v11,v13} to compute isochoric gas cooling functions for 24 metallicities in the range $(10^{-6}-2)~Z_\odot$ (Fig. 1). The grid of metallicities was constructed so that the difference between neighboring metallicities did not exceed a factor of ten, with the cooling rates for neighboring metallicities for any fixed temperature in the interval $10$~K$<T<10^8$~K not exceeding a factor of 1.5. These computed cooling rates are close to the functions obtained in \cite{sd93,gs07} for $T>10^4$~K and the
functions from \cite{sn97} for the interval $10$~K$<T<10^4$~K (Fig. 1).

The velocity of the supernova envelope falls with time, and the gas crossing the front is heated to lower temperatures, $T\sim v^2$. The ionizational and thermal evolution of this gas will differ from the evolution of the gas that is initially heated to several million K (see,
e.g., \cite{v12}). The relaxation time for the ion composition behind the front of a strong shock is shorter than the cooling time, so that the gas behind a shock front with a temperature $T_s \simgt 3\times 10^5$~K rapidly reaches a thermal and ionization state that is the same as if it began to cool from $T=10^8$~K \cite{v12}. Since the shock from the supernova remains fairly strong in our computations, we can use the pre-calculated nonequilibrium cooling functions for gas cooled from $T=10^8$~K without introducing appreciable errors.

%%%%%%%%%%%%%%%%%%%%%%%%%%%%%%%%%%%%%%%%%%%%%%%%%%%%%%%%%%%%%%%%%%%%%
\begin{figure}[!ht]
\center
\includegraphics[width=10cm]{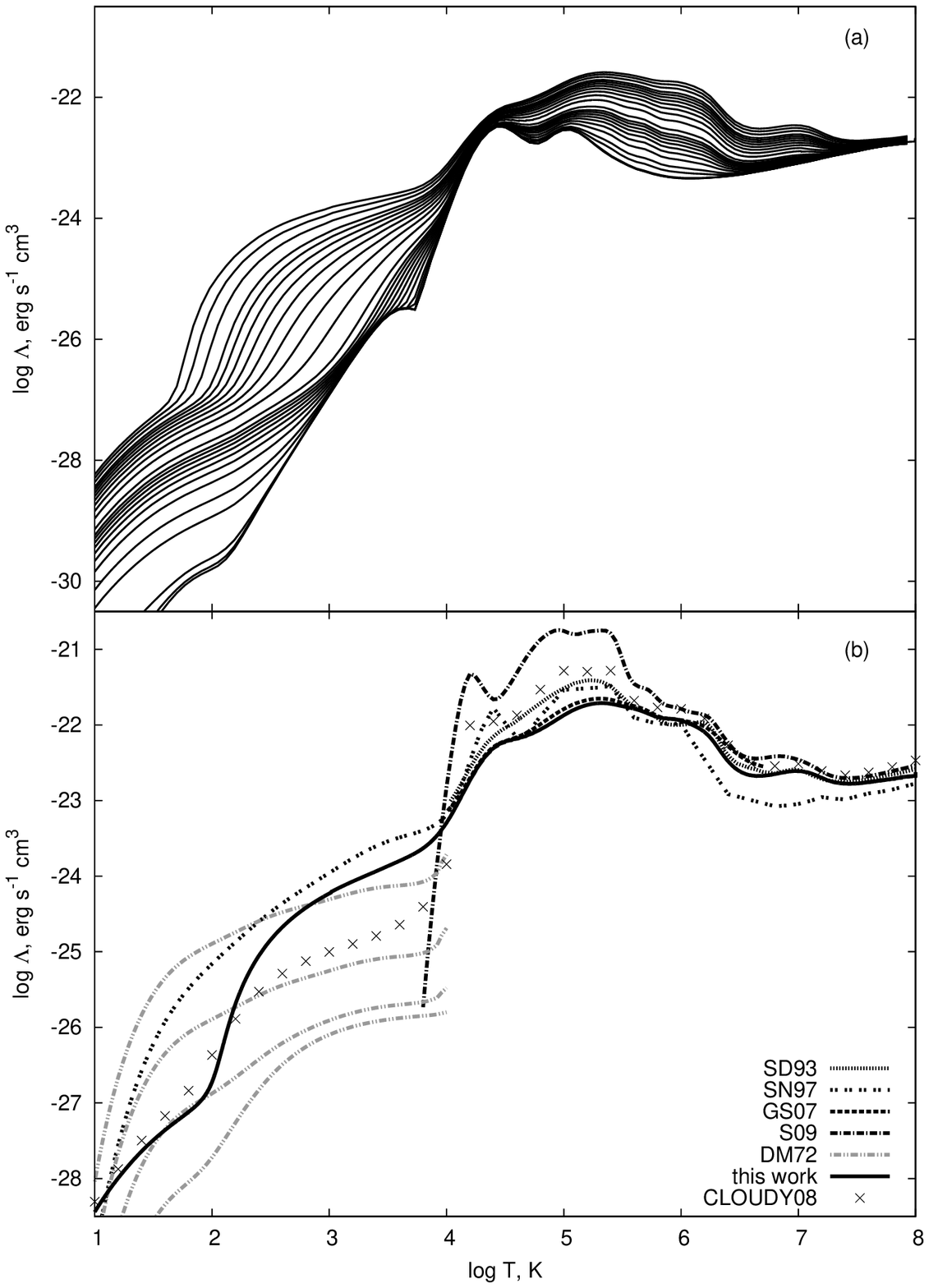}
\caption{
{\it Upper}: gas cooling function for 24 metallicities in the interval $10^{-6}-2~Z_\odot$ used in our study. 
{\it Lower}: cooling function for the solar metallicity in a isochoric gas, calculated using the method of \cite{v13}, 
and the cooling functions obtained in \cite{sd93} -- SD93, a non-equilibrium function; \cite{sn97} -- SN97; \cite{gs07} -- GS07; 
\cite{schure09} -- S09; and \cite{dalgarno72} -- DM72 for relative ionizations $f_i=n_e/n_H = 10^{-4}$, $10^{-3}$, $10^{-2}$ and $0.1$, 
which correspond to the four double-dot-dashed curves, from top to bottom. The cooling function obtained using 
the CLOUDY code (\cite{cloudy}, v.08) is shown by the x’s (the H$_2$ molecule was not included in the eqilibrium computation).
}
\label{figcool}
\end{figure}
%%%%%%%%%%%%%%%%%%%%%%%%%%%%%%%%%%%%%%%%%%%%%%%%%%%%%%%%%%%%%%%%%%%%%

\subsection{Initial Condiions}

We considered the evolution of a supernova remnant with energy $10^{51}$~erg in a cloudy medium with a number density for the inter-cloud gas $n=0.1$~cm$^{-3}$ and a temperature $T=9\times 10^3$~K, corresponding to the values in the interstellar medium of our and other
galaxies. The gas number density and temperature inside the clouds were taken to be $n=15$~cm$^{-3}$ and $T=60$~K. The clouds are in dynamical equilibrium with the ambient medium.

At the initial time, an energy of $10^{51}$~erg, a gas mass $M_g=21~\msun$, and a mass of metals $M_m=2.1~\msun$, corresponding to a supernova mass $M_* = 25~\msun$ \cite{ww95}, are added in the central 3~pc. Thus, the metallicity of the gas in the ejected material (ejecta) is $5~\zsun$. In preliminary computations, the metallicities in the unperturbed cloudy and intercloud media were varied from $[Z/H]=-3$ to $[Z/H]=-1$, with any substantial influence on the results. Therefore, we chose the value $[Z/H]=-3$ to more clearly represent
the redistribution of metals. This enabled us to easily find the corresponding heating rate for which the clouds in their initial state are in equilibrium with the intercloud gas. For the chosen metallicity, the heating rate is $\Gamma = 10^{-25}$~erg/s. This metallicity is appreciably lower than the observed values in the modern-day interstellar media of spiral galaxies, but is quite admissible for some dwarf galaxies (see, e.g., \cite{dwarfs-review}). The use of this metallicity is not of fundamental importance from the point of view of the dynamics of the interaction between a supernova remnant and a cloudy medium, apart from the fact that the total radiative losses in the gas between the shock front and the surface of the ejecta is weakened, so that the remnant remains in the adiabatic phase longer.

The clouds are distributed randomly and uniformly around the supernova. We describe the inhomogeneities in the cloudy medium using the volume filling factor $f$, equal to the ratio of the entire volume of clouds to the volume of the computational region.\footnote{ In an axially symmetrical approach, the clouds form tori, so that the arrangement of the clouds is more rarified than in an intrinsically three-dimensional approach. However, the difference in the mean distance between the clouds in axially symmetrical and fully three-dimensionalmodels is $\simlt 30$\% for the considered range of filling factors, making the axially symmetric approach satisfactory. Note that the fraction of the envelope surface that interacts with clouds—the filling factor—is important for the dynamics of the shock [see formula (\ref{eq-fsurf}) below], which is the same in the two approaches.} We investigated the evolution of a supernova remnant
in cloudy media with filling factors from 0.02 to 0.2. The computational domain was $75\times150$~pc in size with a cell size of 0.075~pc, corresponding to a grid of $1000\times2000$ cells. The cloud radii were normally distributed with a mean value of 1.5~pc and a dispersion
of 1~pc, so that the cloud radii lay mainly in the range from 0.5 to 2.5~pc, and the number of cells corresponding to the cloud radii varied approximately from 6 to 33. This is too few to study the evolution of the disruption of an individual cloud interacting with
a shock \cite{klein94}, but is fully sufficient to investigate the interaction of a supernova envelope with an ensemble of clouds. Computations with grids that are a factor of two and four larger ($2000\times4000$ and $4000\times8000$) showed only unimportant differences in the overall pattern of the interaction and the statistical characteristics of the physical properties of the gas. It was
shown in \cite{klein94} that studies of the dynamics of the disruption of an individual cloud by a shock front require computations with about 100 cells per cloud radius. In computations with grids of $4000\times8000$ cells, the number of cells per cloud radius varies from 25 to 130, which is fully sufficient to trace the evolution of medium-sized and large clouds.

%----------------------- Section 3 -------------------------------

\section{Evolution of a supernova remnant in a cloudy medium}

\noindent

Figure 2 presents the evolution of the density, temperature, and metallicity of the gas after a supernova explosion in a medium with number density $n=0.1$~cm$^{-3}$ and cloud filling factor $f = 0.05$. An obvious characteristic of the evolution of a supernova remnant in a cloudy medium compared to the case of a uniform medium is the strong distortion of the shock front almost immediately after the explosion,
due to clouds located near the site of the supernova. The shape of the supernova envelope remains close to spherical in a uniform medium until hydrodynamical (Rayleigh–Taylor) instability begins to develop, as well as thermal instability in the radiative phase. Thus, an envelope expanding in a uniform medium will be close to spherical approximately until a time corresponding to the cooling time, $t_c \sim 10^6$~yrs. In a cloudy medium, distortion of the envelope arises immediately, as soon as the front reaches the clouds located closest to the supernova site.

On the other hand, the interaction of the shock and clouds can lead to a relatively early onset of the radiative phase in the corresponding part of the supernova envelope. Moreover, in the case of a supernova expanding in a cloudy medium, the shock front tears apart dense and cool clouds located close to the explosion site, nearly fully disrupting them; the material from these clouds enters the hot, rarified cavity, increasing the density and the loss of thermal energy by the hot gas in this cavity.

Thus, in a supernova explosion in a cloudy medium, the radiative phase does not begin simultaneously in the entire envelope (see the definition of a supernova envelope and the hot cavity in the Introduction), as can be seen in the plot for time $t = 1.3\times 10^5$~yrs~$< t_c$ in Fig. 2. The higher the volume filling factor $f$, the larger the fraction of the envelope that interacts with clouds, and the earlier the envelope makes the transition to the radiative phase. The surface factor -- the fraction of the envelope surface that interacts
with clouds -- is approximately $f_s \simeq fR_s/4a$ ($R_s \simeq (Et^2/\rho)^{1/5}$ is the radius of the supernova envelope and a the mean radius of the clouds), or in normalized form,
\be
 f_s\simeq 0.7 \ \left( {f \over 0.05}\right)
                 \left( {E \over 10^{51}~\rm erg }  { 0.1~{\rm cm^{-3} } \over n }\right)^{1/5}
                 \left( {t \over 10^5~{\rm yr}} \right)^{2/5} \left( {1~{\rm pc} \over a}\right).
\label{eq-fsurf}
\ee
With a volume filling factor $f = 0.05$, nearly the entire envelope surface is able to interact with clouds by $t = 2.3\times 10^5$~yrs; i.e., the entire envelope makes the transition to the radiatie phase by this time. This shows thatmass loading can substantially change the dynamics of the entire supernova remnant.

%%%%%%%%%%%%%%%%%%%%%%%%%%%%%%%%%%%%%%%%%%%%%%%%%%%%%%%%%%%%%%%%%%%%%
\begin{figure}[!ht]
\center
\includegraphics[width=14.5cm]{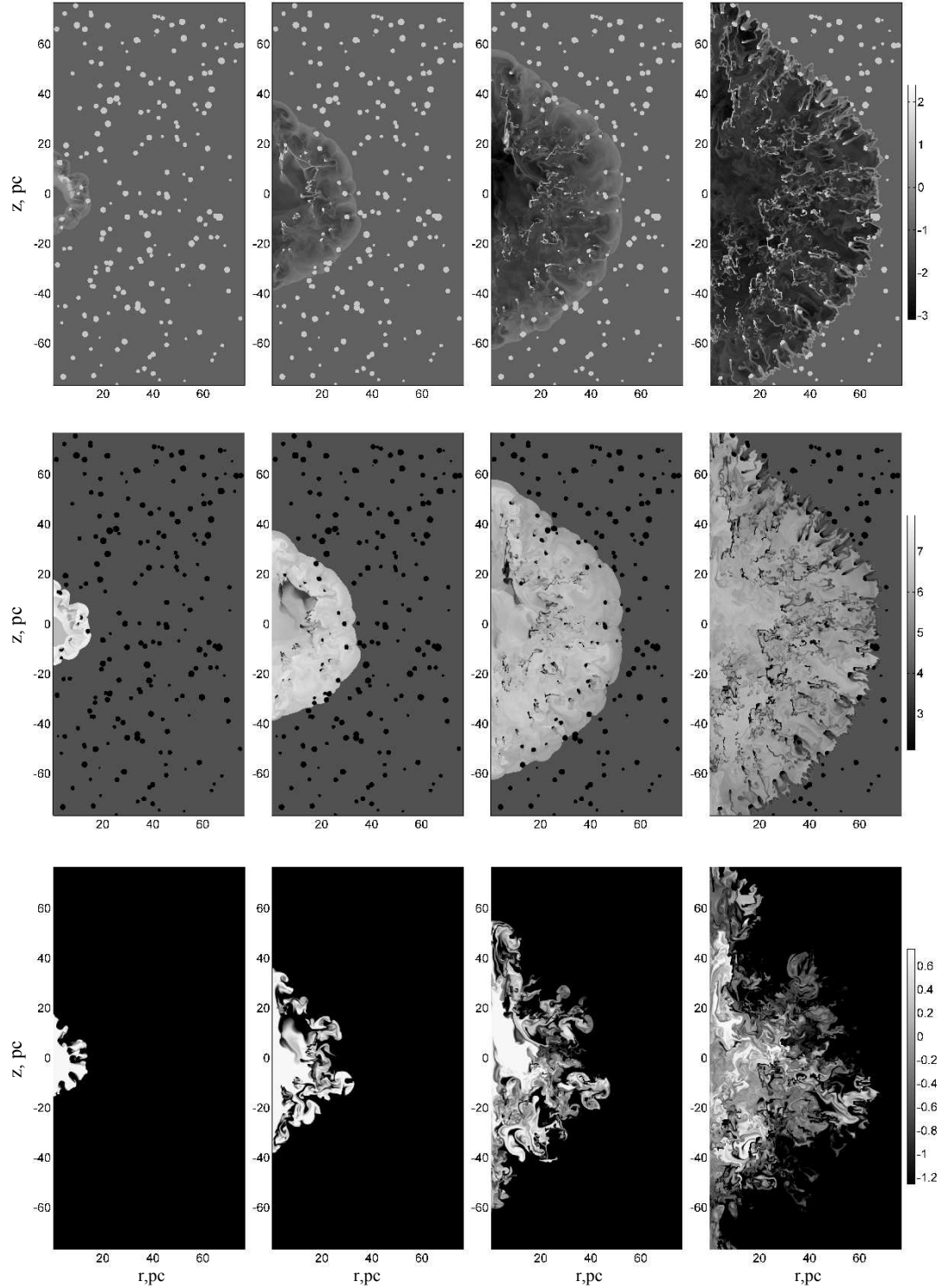}
\caption{
Maps of the distributions of the density (upper row), temperature (middle row), and metallicity (lower row) of the gas
at times $t = 8.8\times10^3, 4.4\times10^4, 1.3\times10^5, 3.5\times10^5$~yrs after a supernova explosion in a medium with filling factor
$f=0.05$.
}
\label{evol}
\end{figure}
%%%%%%%%%%%%%%%%%%%%%%%%%%%%%%%%%%%%%%%%%%%%%%%%%%%%%%%%%%%%%%%%%%%%%

Let us consider the evolution of a supernova remnant in more detail. The shock front is weakened and distorted by clouds located nearby ($r\simlt 10$~pc); this is clearly visible in a comparison of the density distributions in the upper and low half-planes (relative to the coordinate origin). Clouds located at radii $r\simlt 10$~pc are nearly fully disrupted by $t\sim (0.5-1) \times 10^5$~yrs (Fig. 2). Clouds that are at larger radii ($r\sim 10-20$~pc) lose a substantial fraction of their mass during the passage of the shock, but the most massive of them survive and are accelerated. These clouds undergo further disruption as they move in the hot, rarified gas. When it reaches $r\simgt 20$~pc, the shock front is sufficiently weakened that a large fraction of the clouds passing through it survive, although their destruction due to Kelvin–Helmholtz instability is still rather efficient. After $t \sim 2\times 10^5$~yrs, the supernova shock is no longer capable of accelerating the clouds. At the end time of the computations, $t = 3.5\times 10^5$~yrs, clouds at distances $r\sim 20-30$~pc have been transformed into extended objects, while those located at radii $r\simgt 50$~pc have been preserved after their interaction with the supernova envelope. The outer parts of these clouds\footnote{The dynamical evolution of the clouds was described using
computations on a $4000\times8000$ grid, which corresponds to 25–130 cells per cloud radius, and fully corresponds to the recommendations of \cite{klein94} for studies of the disruption of an individual cloud.}are eroded by the development of Kelvin–Helmholtz instability, and they are followed by tails resembling cometary tails; however, the largest retain a close-to-spherical shape.

The pressure inside fragments and the extended tails is appreciably lower than in the hot cavity. This means that these structures are compressed by the external pressure, while being gradually disrupted by the development of Kelvin–Helmholtz instability over a time from tens to hundreds of thousands of years. Only at late times $t \simgt 2.5\times 10^5$~yrs, when the supernova envelope has cooled and slowed to low Mach numbers, does the pressure in the clouds crossing the front become close to the pressure in the outer parts of the supernova envelope. Since the gas inside fragments has close to the initial temperature of the cloud (60~K), and the tails extending behind the clouds also have temperatures lower than the ambient temperature, a very non-uniform temperature distribution forms inside the cavity (Fig. 2). Thus, the interaction of a supernova remnant with a cloudy medium is characterized by appreciable density and temperature inhomogeneity of the gas, both in the supernova envelope and in the hot cavity.

The supernova envelope interacts with numerous clouds, disrupting them and transferring part of its kinetic energy to them. An appreciable fraction of this energy goes into the energy of chaotic motions of fragments surrounded by the hot gas of the cavity. Many cloud fragments that have collided with each other, forming secondary shocks and acoustic waves, new clouds, and vortices, and thereby transforming kinetic into thermal energy, can be noted inside the cavity in Fig. 2. At early stages of the evolution of the supernova envelope ($t\sim (0.5-1)\times 10^5$~yrs), material from disrupted clouds enters the hot envelope, increasing the mean density of the gas, and possibly leading to an early onset for the cooling of the cavity.

Metals that were initially contained in a small volume are spread behind the shock. The gas with high metallicity is hottest and most rarified inside the cavity, and its cooling time appreciably exceeds the age of the envelope. In the case of a supernova in a uniform medium, favorable conditions for the development of hydrodynamical instabilities arise at the boundary of the cavity and envelope starting from the onset of the radiative phase; i.e., after appreciable cooling of the gas in the envelope, accompanied by a drop in the pressure in the cavity. Due to the development of instability, the unenriched, dense gas of the envelope mixes with the high-metallicity, rarified gas in the cavity; the efficiency of this mixing is obviously low (see, e.g., \cite{wang01,v08,mix12}).

The global shock front is appreciably distorted in the interaction with the cloudy medium, and the clouds behind the front are partially or fully disrupted. This leads to the formation already in the adiabatic phase of the supernova expansion of numerous local shocks and dense, chaotically distributed fragments in the hot, metal-enriched cavity (Fig. 2). The energy of local shocks and the disruption of fragments due to hydrodynamical instability facilitate an increase in the mass of enriched gas. Consequently, the efficiency of the mixing of metals inside the envelope also changes. A detailed statistical analysis of the distribution of metals and the efficiency of mixing will be carried out in a separate study.

%%%%%%%%%%%%%%%%%%%%%%%%%%%%%%%%%%%%%%%%%%%%%%%%%%%%%%%%%%%%%%%%%%%%%
\begin{figure}[!ht]
\center
\includegraphics[width=12cm]{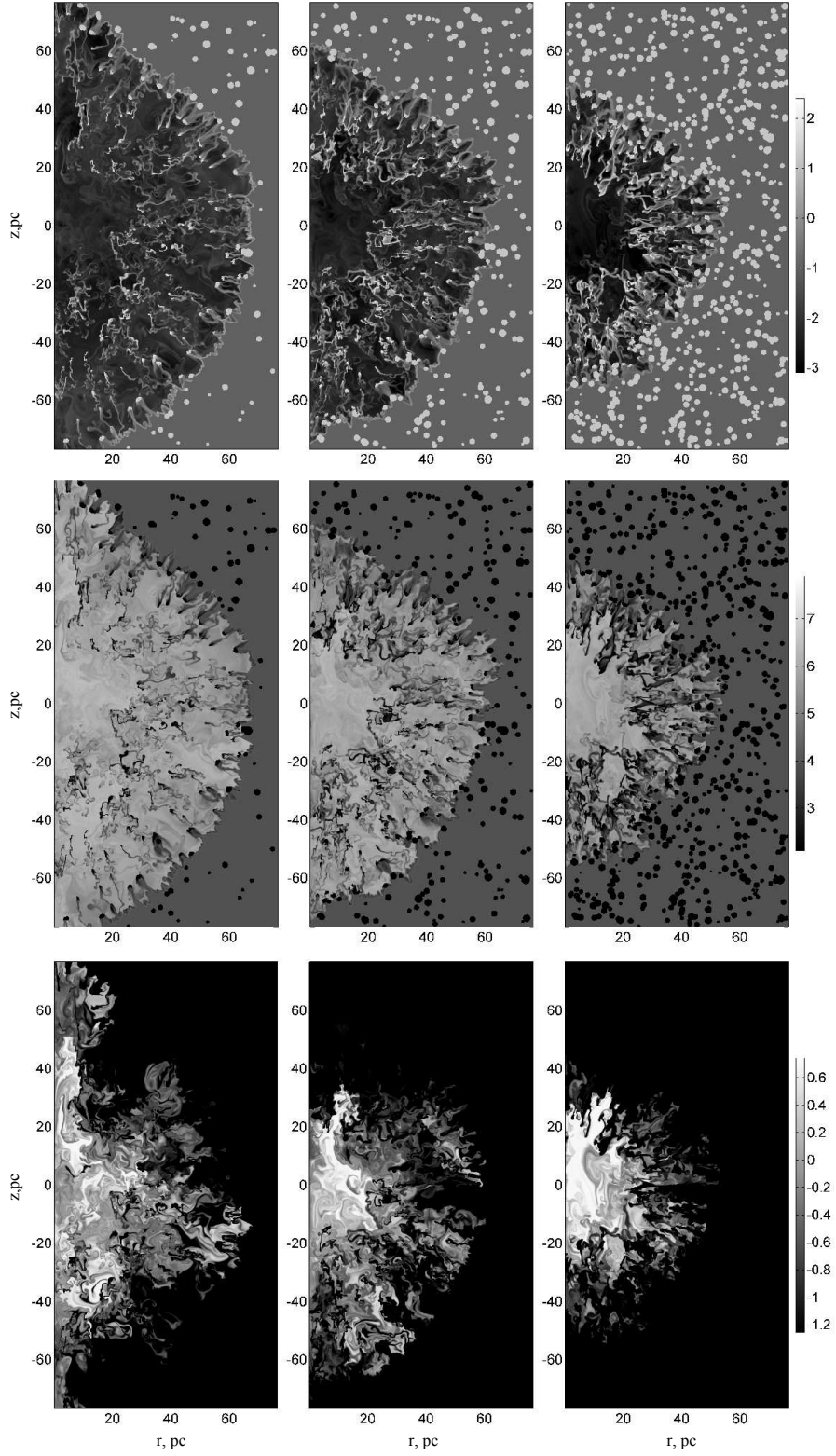}
\caption{
Maps of the distributions of the density (upper row), temperature (middle row), and metallicity (lower row) of the gas at
the final time $t = 3.5\times 10^5$~yrs after supernova explosions in media with filling factors $f=0.05, 0.1, 0.2$ (from left to right).
}
\label{evolfc}
\end{figure}
%%%%%%%%%%%%%%%%%%%%%%%%%%%%%%%%%%%%%%%%%%%%%%%%%%%%%%%%%%%%%%%%%%%%%

%----------------------- Section 4 -------------------------------
\section{Dependence on the filling factor}

\noindent

As was noted earlier, a large density of the gas in the clouds facilitates an earlier onset for the radiative phase in the parts of the supernova envelope with which they interact. The distortion of the front increases with the volume filling factor. Figure 3 presents the distributions of the density, temperature, and metallicity for media with filling factors $f=0.05, 0.1, 0.2$ at time $t=3.5\times 10^5$~yrs after the supernova. These clearly show that the shape of the supernova shock front becomes more complex as the filling factor increases, while the mean radius of the supernova envelope decreases. Figure 4 presents the evolution of the mean radius of the supernova envelope in media with filling factors $f = 0.05, 0.1, 0.2$. The difference in the mean size of the envelope is manifest nearly at the very beginning  of the evolution, and reaches approximately factors of 1.3 and 1.5 by the end of the computation for volume filling factors of 0.05 and 0.2, respectively.

%%%%%%%%%%%%%%%%%%%%%%%%%%%%%%%%%%%%%%%%%%%%%%%%%%%%%%%%%%%%%%%%%%%%%
\begin{figure}[!ht]
\center
\includegraphics[width=10cm]{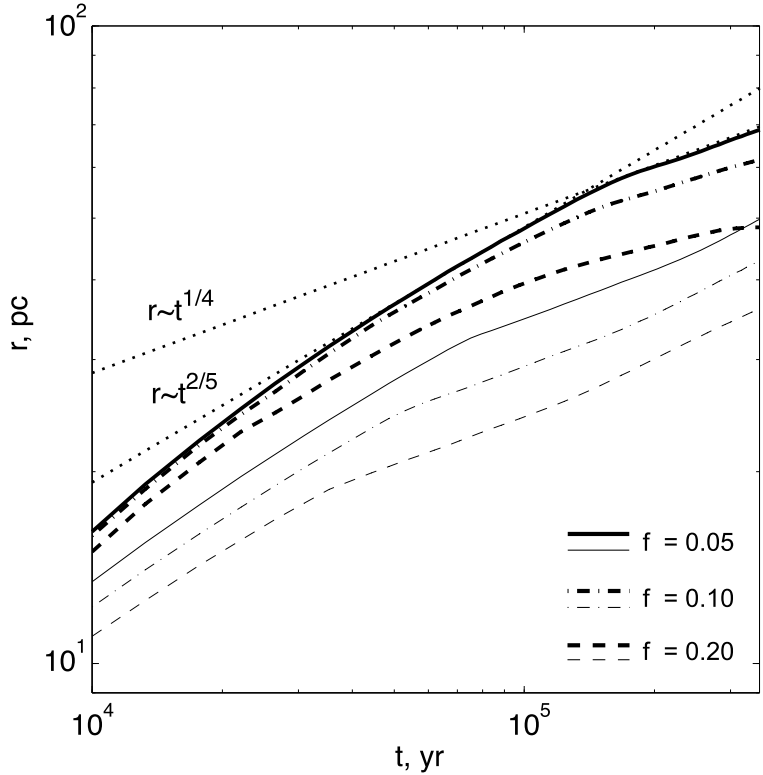}
\caption{
Mean radius of the supernova envelope in the case of explosions in cloudy media with filling factors $f=0.05, 0.1, 0.2$
(thick dash–dot, and dashed curves, respectively) and in media with uniformly distributed cloudy-medium gas (corresponding
thin curves). The dotted curves show the dependences $r\sim t^{2/5}$ and $r\sim t^{1/4}$, corresponding to the adiabatic and radiative
phases of the supernova.
}
\label{evolrad}
\end{figure}
%%%%%%%%%%%%%%%%%%%%%%%%%%%%%%%%%%%%%%%%%%%%%%%%%%%%%%%%%%%%%%%%%%%%%

The adiabatic phase ($r\sim t^{2/5}$) and radiative phase ($r\sim t^{1/4}$) in the evolution of the supernova envelope can clearly be distinguished (Fig. 4). As was noted above, increasing the filling factor accelerates the onset of the radiative phase roughly as $f^{-1/2}$: the influence of the clouds on the onset of the radiative phase is appreciable for $f = 0.05$, when the transition to Oort expansion occurs at $t\sim 1.5\times 10^5$~yrs; when $f\sim 0.1$ and $f\sim 0.2$, cooling becomes important by $t\sim 10^5$~yrs and $t\sim 0.7\times 10^5$~yrs, respectively. Subsequent increase in the filling factor leads not only to an even earlier onset of the radiative phase, but also to variation in the expansion of the envelope, compared to the Oort law $r\sim t^{1/4}$; for $f\sim 0.2$ these differences
are appreciable at times $t\simgt 1.5\times 10^5$~yrs. The envelope is rapidly decelerated, and its size remains nearly constant after $t\sim 2.7\times 10^5$~yrs, which corresponds to the total loss of the kinetic energy of the global shock front and the transfer of the momentum of the envelope to the medium.

To illustrate the influence of inhomogeneity of the interstellar medium on the dynamics of the remnant, we homogenized the medium, by uniformly distributing the mass of the cloudy phase in a diffuse medium with a constant density and the same total mass of gas. In this case, the density of the homogeneous diffuse medium increased by a factor of $\chi f$ compared to the density of the background intercloud gas of the initially inhomogeneous medium, where $\chi = 140$ is the enhancement of the density in the cloudy phase over the background value. This means that the radius of the remnant at the adiabatic phase will be smaller by a factor of $\sim (140 f)^{-2/5}$, compared to the case of evolution in a cloudy medium with surface factor $f_s \ll 1$. The thin curves in Fig. 4 show the radius of the supernova envelope following its explosion in such a homogenized medium. Note that the envelope is smaller, and the transition to the radiative phase earlier compared to the case of a cloudy medium, in spite of the fact that the mean density in the homogenized medium is lower than the density in the clouds.

The growth in the surface filling factor (\ref{eq-fsurf}) as the spherical shock propagates in a cloudy medium is proportional to the radius of the shock. However, since the cross section of an individual cloud is substantially less than the total surface of the supernova envelope, the shock passes around the clouds, losing only an insignificant fraction of its energy and momentum. When the mass of the clouds is redistributed homogeneously, the supernova envelope propagates in a medium with an enhanced density from the very earliest times, leading to an early onset of the radiative phase. Thus, an inhomogeneous distribution of thematter ahead of the shock front appreciably influences the onset of the radiative phase and the energy-loss rate by the supernova remnant, and can lead to appreciable differences from the case when the shock expands in a homogeneous medium \cite{chevalier74,cox72}.

Figure 5 depicts the evolution of the thermal and kinetic energy of a supernova remnant\footnote{The thermal (kinetic) energy of the remnant is the sum of the thermal (kinetic, including the chaotic motions of the gas) energy of the gas contained inside the sphere of radius $r$, depicted in Fig. 4.} in a homogenized medium in which the gas of the cloudy medium has been uniformly redistributed (thin curves) and in a cloudy medium (thick curves). We can clearly see that the thermal energy is constant up to the onset of the radiative phase in the case of a uniform medium. The thermal energy then decreases with time as $\sim t^{-1.2}$ \cite{cioffi88,thornton98} right up to the onset of the development of hydrodynamical instability in the supernova envelope, i.e., up to the disruption of the envelope, after which a thermal-energy loss rate $\sim t^{-2.3}$ us established in the envelope (thin curves in Fig. 5).

%%%%%%%%%%%%%%%%%%%%%%%%%%%%%%%%%%%%%%%%%%%%%%%%%%%%%%%%%%%%%%%%%%%%%
\begin{figure}[!ht]
\center
\includegraphics[width=10cm]{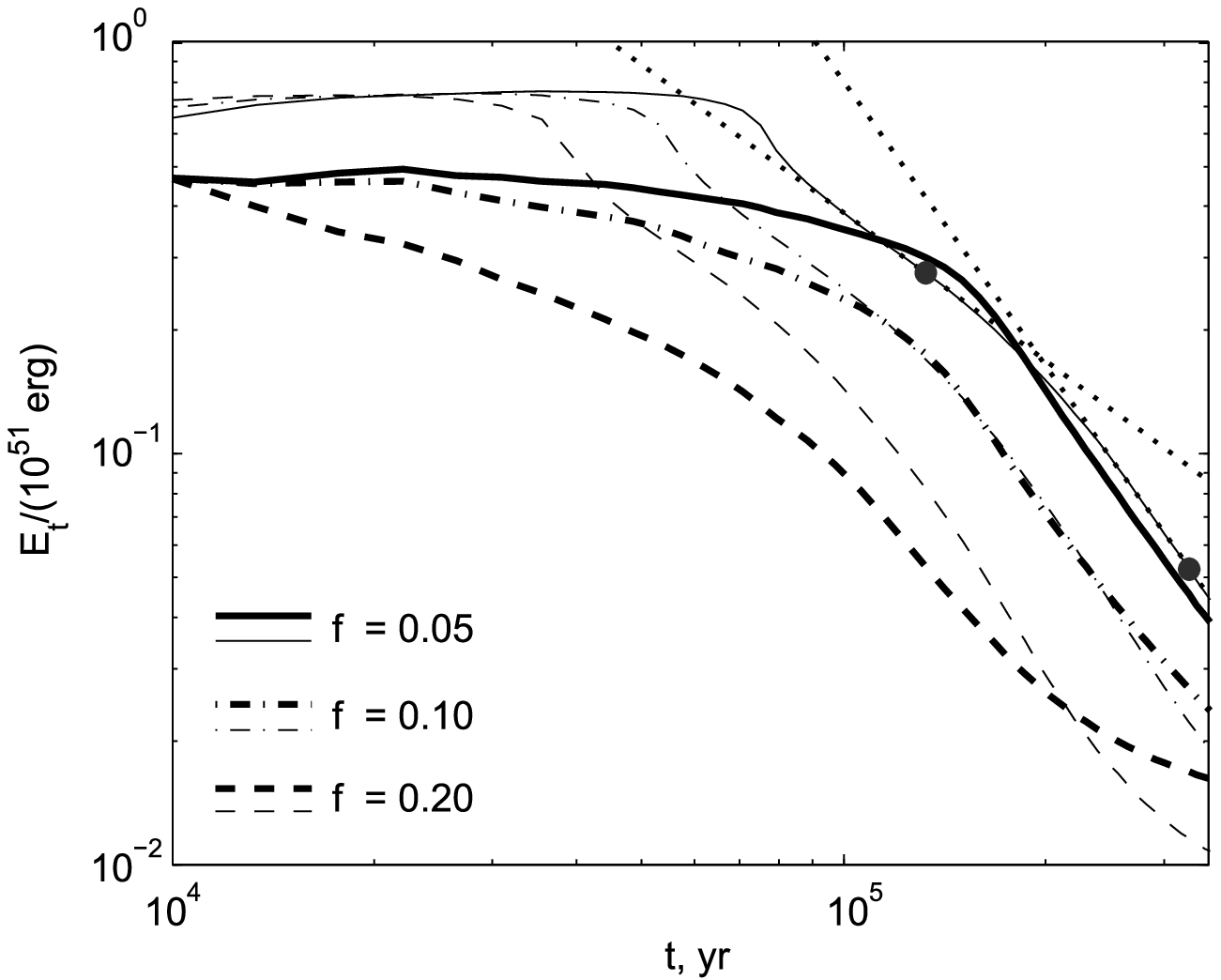}
\includegraphics[width=10cm]{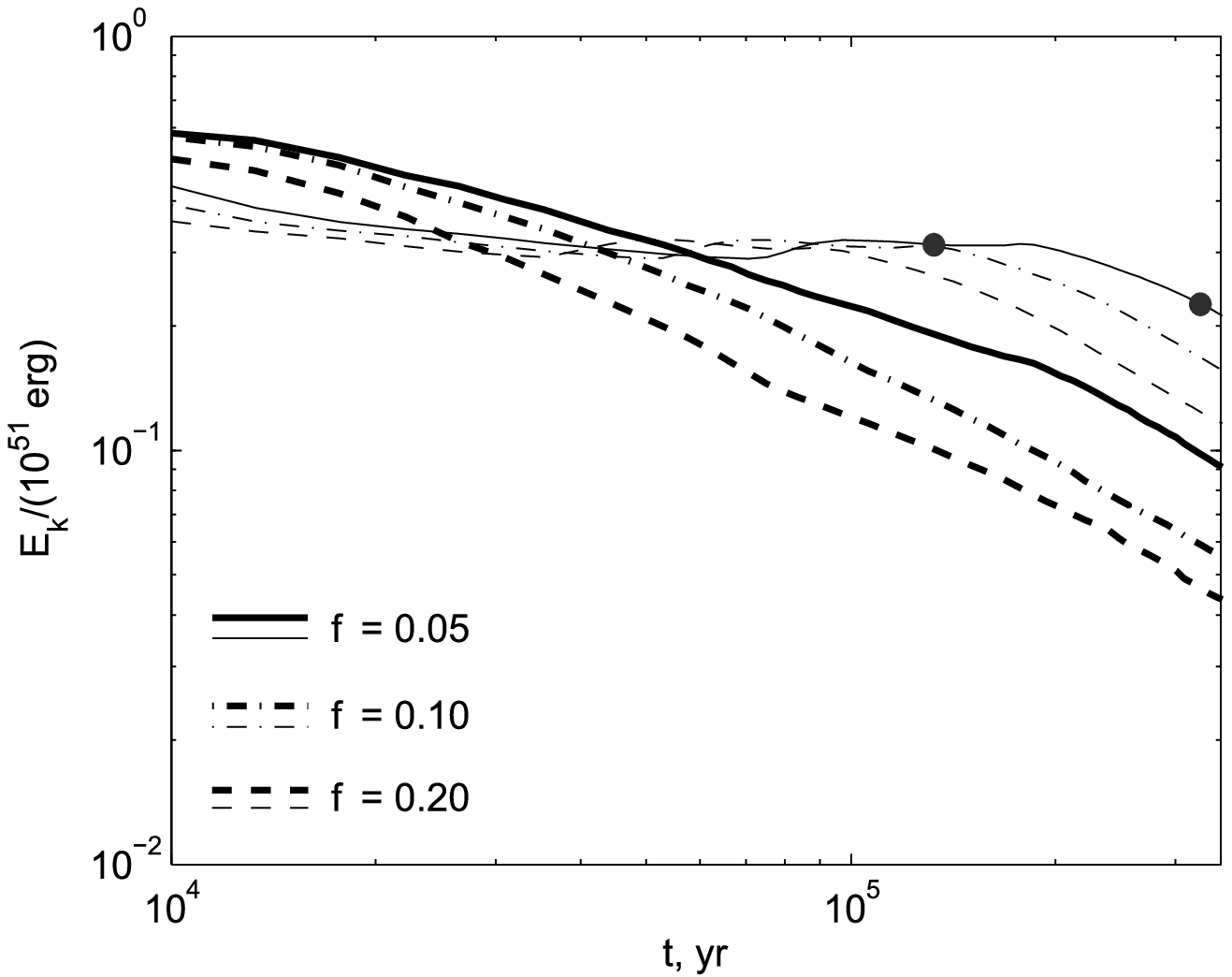}
\caption{
Thermal (upper) and kinetic (lower) energy of the supernova envelope in cloudy media (thick curves) and in homogenized media in which the mass of the clouds is uniformly distributed (thin curves). The models for volume filling factors $f=0.05, 0.1, 0.2$ are shown by the solid, dashed, and dot-dashed curves, respectively. The dotted curves show the dependences $E \sim t^{-1.2}$ and $E\sim t^{-2.3}$. The filled circles mark times $t = 1.3\times10^5$ and $3.23\times10^5$~yrs.
}
\label{evol-en}
\end{figure}
%%%%%%%%%%%%%%%%%%%%%%%%%%%%%%%%%%%%%%%%%%%%%%%%%%%%%%%%%%%%%%%%%%%%%

%%%%%%%%%%%%%%%%%%%%%%%%%%%%%%%%%%%%%%%%%%%%%%%%%%%%%%%%%%%%%%%%%%%%%
\begin{figure}[!ht]
\center
\includegraphics[width=10.5cm]{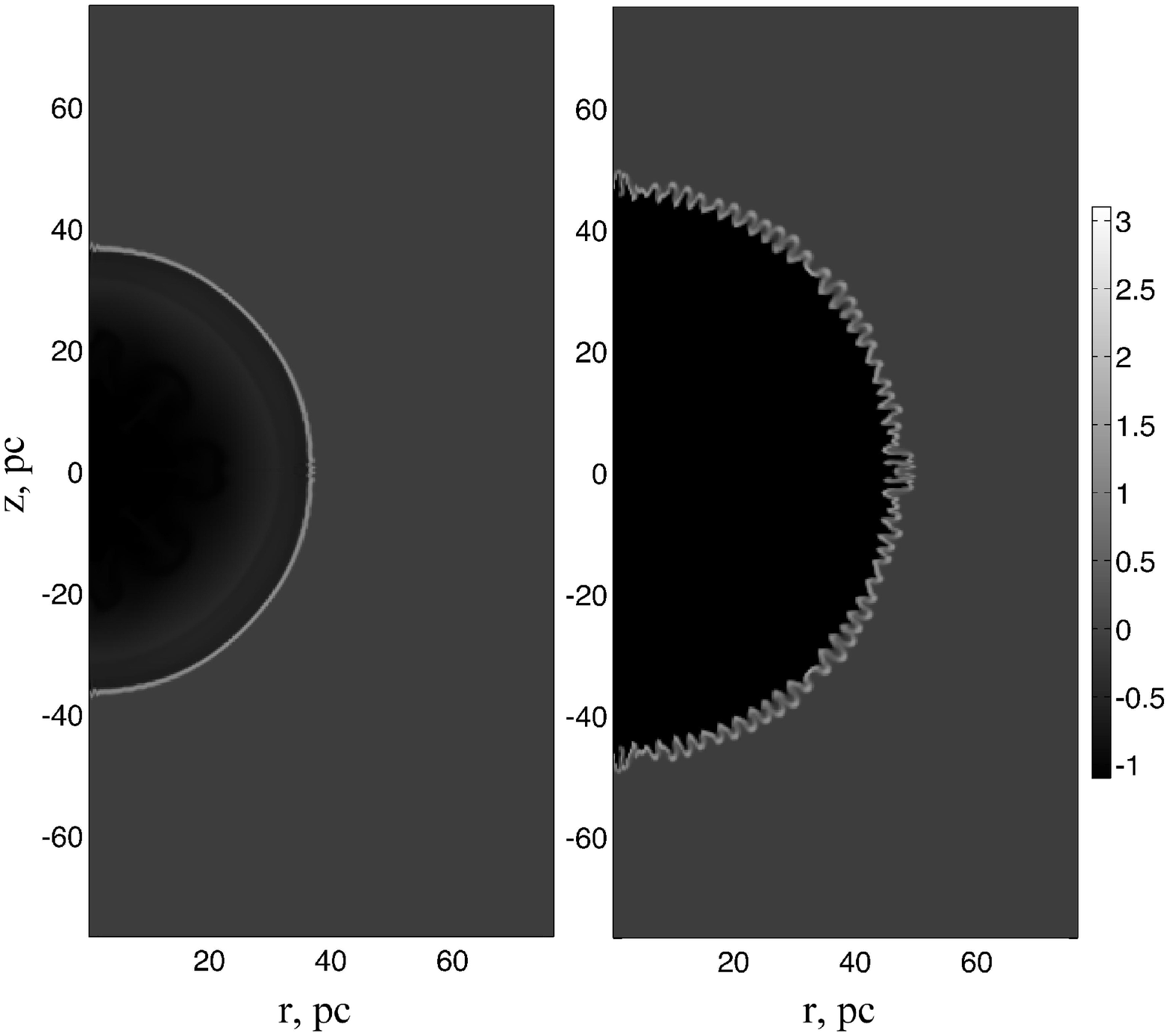}
\caption{
Maps of the density distribution of the gas at times $t = 1.3\times10^5$~yrs (left) and $t = 3.2\times10^5$~yrs (right) after a supernova
in a uniform medium with density $\chi f\rho$ for $f=0.05$.
}
\label{evolsnhomo}
\end{figure}
%%%%%%%%%%%%%%%%%%%%%%%%%%%%%%%%%%%%%%%%%%%%%%%%%%%%%%%%%%%%%%%%%%%%%

As an example, Fig. 6 presents distributions of the gas density at times $t = 1.3\times10^5$ and $3.2\times10^5$~yrs after a supernova explosion in a uniform medium with density $\chi f\rho$ for $f=0.05$. These times are marked by filled circles in Fig. 5. We can clearly see that the envelope is in the radiative phase at $t = 1.3\times10^5$~yrs, but no instability is obvious. The growth of instability in the envelope and the transition to the non-linear stage has begun at $t \sim 2\times10^5$~yrs, and the fragmentation of the envelope is fairly clearly manifest by $t \simgt 3\times10^5$~yrs (Fig. 6). In the development of instability in the envelope, its surface and thickness grow, the gas expands adiabatically, and the thermal energy falls. The appearance of instability is accompanied by the loss of strictly radial motion of the envelope’s expansion and the appearance of tangential gas flows in growing fragments of the envelope, which can subsequently collide with each other. As a consequence, the kinetic energy begins to decrease at the onset of the supernova-envelope’s disruption (thin curves in the lower panel of Fig. 5).

In a cloudy medium, the onset of the radiative phase is not simultaneous in all parts of the supernova envelope: it begins earlier in those parts of the envelope that interact with clouds. Therefore, in the inhomogeneous case, the thermal energy falls beginning from the earliest stages of the envelope expansion, and the energy losses grow with increasing filling factor (thick curves in Fig. 5). After the onset of cooling in the entire envelope, including in those regions that interact only with the inter-cloud gas, the thermal-energy loss rate is $\sim t^{-2.3}$; i.e., the same as after the onset of the envelope disruption during the propagation of the supernova remnant in a uniform medium. The kinetic energy of the remnant in a cloudy medium begins to decrease immediately after the first interaction with the cloudy medium. When the filling factor is small, $f\simlt 0.05$, the kinetic-energy loss rate is enhanced after the onset of cooling in the entire envelope, and depends on time as $\sim t^{-0.9}$ (see, e.g., the solid thick curve in Fig. 5 after $t\sim 2\times 10^5$~yrs). The development of instability in the envelope (and generally in the supernova remnant) begins appreciably earlier in the case of a high filling factor, and the kinetic energy falls as $\sim t^{-0.9}$ beginning from a time $t\sim 10^4$~yrs.

The more efficient energy losses for an envelope  propagating in a cloudy medium with a high filling factor are manifest through a decrease in the volume occupied by the hot gas: when the filling factor is increased from $f = 0.05$ to $f = 0.2$, the volume of gas with $T>10^6$~K and $T=10^5-10^6$~K decreases by roughly a factor of ten (thick curves in Fig. 7). In spite of the fact that the mass of gas in the cloudy medium grows proportional to the filling factor, the mass and volume of the hot ($T>10^6$~K) and warm ($T=10^5-10^6$~K) gas rapidly falls as the filling factor is increased from 0.05 to 0.2: the decrease in themass of the hot component can reach factors of 40–200 for a given time (thick curves in Fig. 7).

The mass fraction of the hot phase is lower for a supernova in a cloudy medium than in a uniform gas with the mean density. This is due, first, to the fact that, in the case of the cloudy medium, the shock propagates through inter-cloud gas with an appreciably lower density than the density of a uniform gas in which the clouds have been smeared throughout, and second, to the presence of cool and dense clouds around which the shock passes, leading to its partial disruption. The temperature of the gas in fragments is appreciably lower than for the interaction of the shock with the less dense inter-cloud gas. The cloud fragments are gradually vaporized, supplementing the diffuse ($10^4$K$<T<10^5$K) and warm ($10^5$K$<T<10^6$K) phases of the supernova remnant, so that the mass fractions of these phases increase with time.

As was already noted above, the presence of a cloudy phase leads to a transfer of momentum from the supernova envelope by fragments and local shocks and vortical flows that form after the disruption of the clouds. The efficiency with which the shock front is decelerated and local shocks form can be estimated from the dependence of the kinetic energy of the layer of gas with radius $r$, $E_k(r)$, and the momentum flux through the outer surface of this layer, $J^p(r)$. Figure 8 depicts $E_k$ and $J^p$ for filling factors $f = 0.05, 0.1, 0.2$ at a time $t = 3.5\times 10^5$~yrs after the supernova. With increasing filling factor, the region with the maximum kinetic energy shifts inward in the envelope. The interaction of the global shock with the clouds weakens the shock and turbularizes the flow behind the front, where a layer of gas from the disrupted clouds forms. As the filling factor increases, the thicknes of this cloudy layer also increases, reaching half the radius of the remnant, on average. This is accompanied by a redistribution of the kinetic energy between the global shock front and numerous vortices that form during the disruption of the clouds. 

Note that a large fraction of the kinetic energy of the remnant is contained in fragments of disrupted clouds. In the case of a supernova that explodes in a medium with a low filling factor, the kinetic energy is mainly concentrated in the vicinity of the global shock front. We can clearly see that, when $f = 0.2$, the bulk of the gas kinetic energy is concentrated in the inner region $r\sim 25-40$~pc, while the mean radius of the envelope reaches nearly 50~pc (Fig. 4). The momentum flux of the gas behaves similarly. Note that the gaseous layer containing the bulk of the kinetic energy and momentum of the remnant essentially traps the hot gas of the cavity in the central region of the supernova remnant (Fig. 3).

Consequently, a growth in the volume filling factor leads to a substantial redistribution of the energy and momentum of the supernova envelope. The layer of cloud fragments that contains the bulk of the energy and momentum of the remnant has a very irregular structure that resembles a net, with more dense and cool fragments located at the nodes of the net. Subsequently, these fragments continue to be disrupted, with the momentum transferred to smaller and smaller scales.

A global shock front propagating in a cloudy medium separates into numerous individual local shocks (Fig. 3), losing a substantial fraction of its energy andmomentumin the process. This weakened shock ceases to efficiently disrupt clouds passing through it. However, since these clouds are subject to the influence of the shock (even if it is relatively weak), they are perturbed from a state of dynamical equilibrium with the surrounding gas, begin to dissipate weakly, and add a small fraction of their mass to the general flow.\footnote{An investigation of the mass-loss rates of the clouds lies beyond the framework of our current study, and will be considered separately.} As a result, a flow of gas having the properties of a mass-loaded flow forms \cite{dyson88}. When $f=0.1$, this picture is achieved by $t\sim 2\times 10^5$~yrs, while this occurs earlier when $f = 0.2$, by $t \sim 1.3\times 10^5$~yrs. Subsequently, the global shock front decelerates, stopping when it has transferred nearly all of its momentum to clouds. When $f = 0.2$, the mean radius of the envelope ceases to increase beginning from $t\sim 3\times 10^5$~yrs (Fig. 4), so that the flow behind the front essentially stabilizes. In the case of a small volume filling factor $f = 0.1$, deviations from the expansion law for the radiative phase, $r\sim t^{1/4}$, become appreciable starting from $t \sim 2\times 10^5$~yrs, but the effects of mass loading are not as strong, and the envelope continues to expand up to the end of the computations.

With the minimum filling factor $f = 0.05$, the shock remains strong at large distances from the site of the supernova explosion, and the clouds behind the front are rapidly disrupted, mainly due to the action of Kelvin–Helmholtz instability. The flow makes a transition to a turbulent regime with an active dynamical exchange of mass and momentum between the clouds and the inter-cloud gas -— a phase of mixing begins \cite{polud02,klein94,mat07}. Thus, when the filling factor is increased, the influence of the supernova shock on the efficiency with which the clouds are disrupted falls appreciably as the radius of the supernova envelope increases. However, due to the large number of clouds, their fragments and tails interact with each other, strengthening their disruption. Although the gas density in these extended, transient structures remains fairly high, the mass of the fragments gradually decreases, and the gas becomes part of the inter-cloud medium. Therefore, the gas behind a supernova shock front propagating through a cloudy medium with volume filling factor $f\sim 0.1-0.2$ exhibits characteristic properties of a flow with mass loading, even for clouds with high density contrasts $\chi \sim 140$. Upon further increase in the filling factor, the supernova shock front rapidly decelerates and the clouds are weakly disrupted; i.e., a small fraction of the mass of the clouds enters the inter-cloud medium. The properties of a mass-loaded flow disappear, and the phase of mixing of the cloud material does not appear \cite{polud02}.

%%%%%%%%%%%%%%%%%%%%%%%%%%%%%%%%%%%%%%%%%%%%%%%%%%%%%%%%%%%%%%%%%%%%%
\begin{figure}[!ht]
\center
\includegraphics[width=7.5cm]{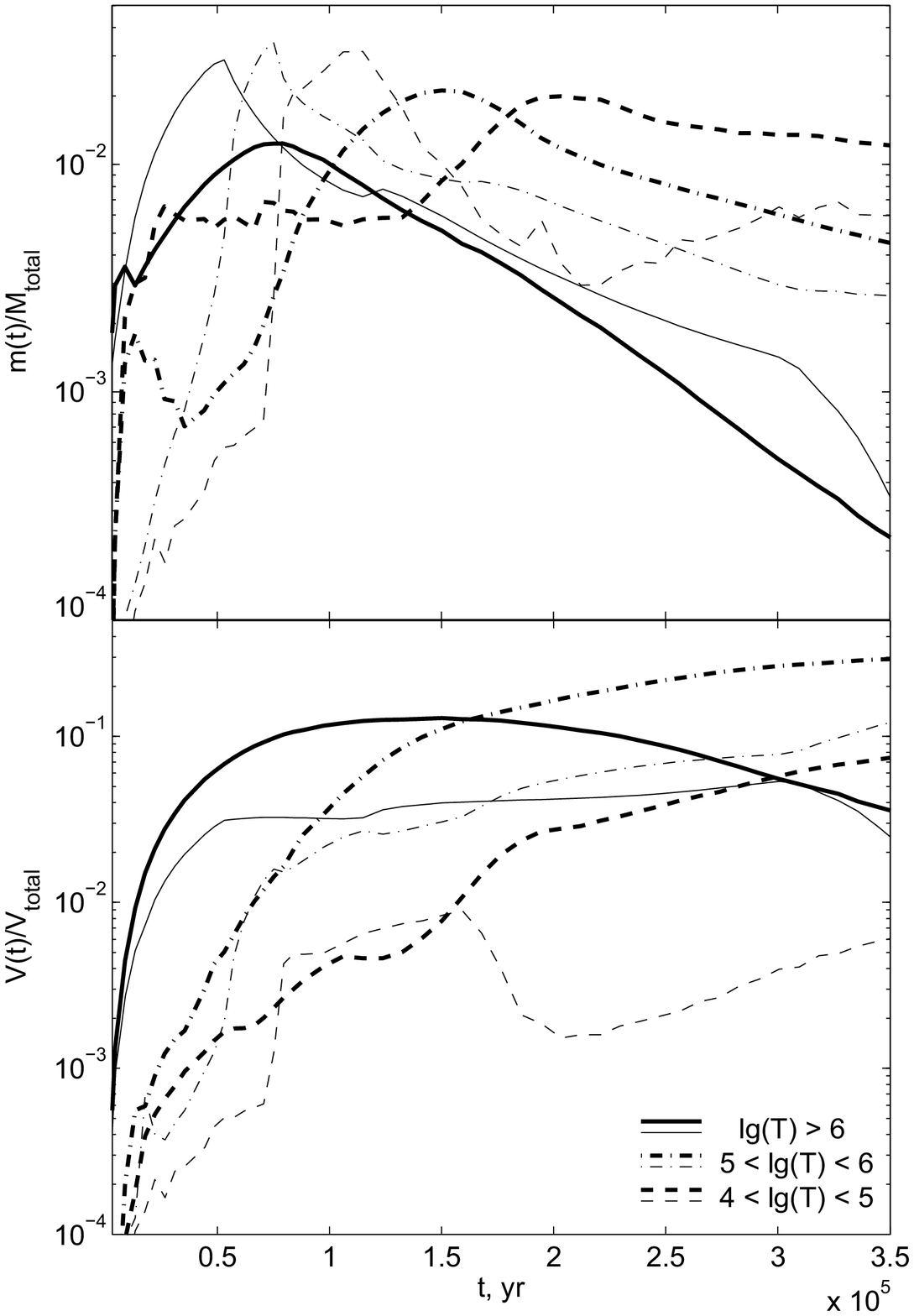}
\includegraphics[width=7.5cm]{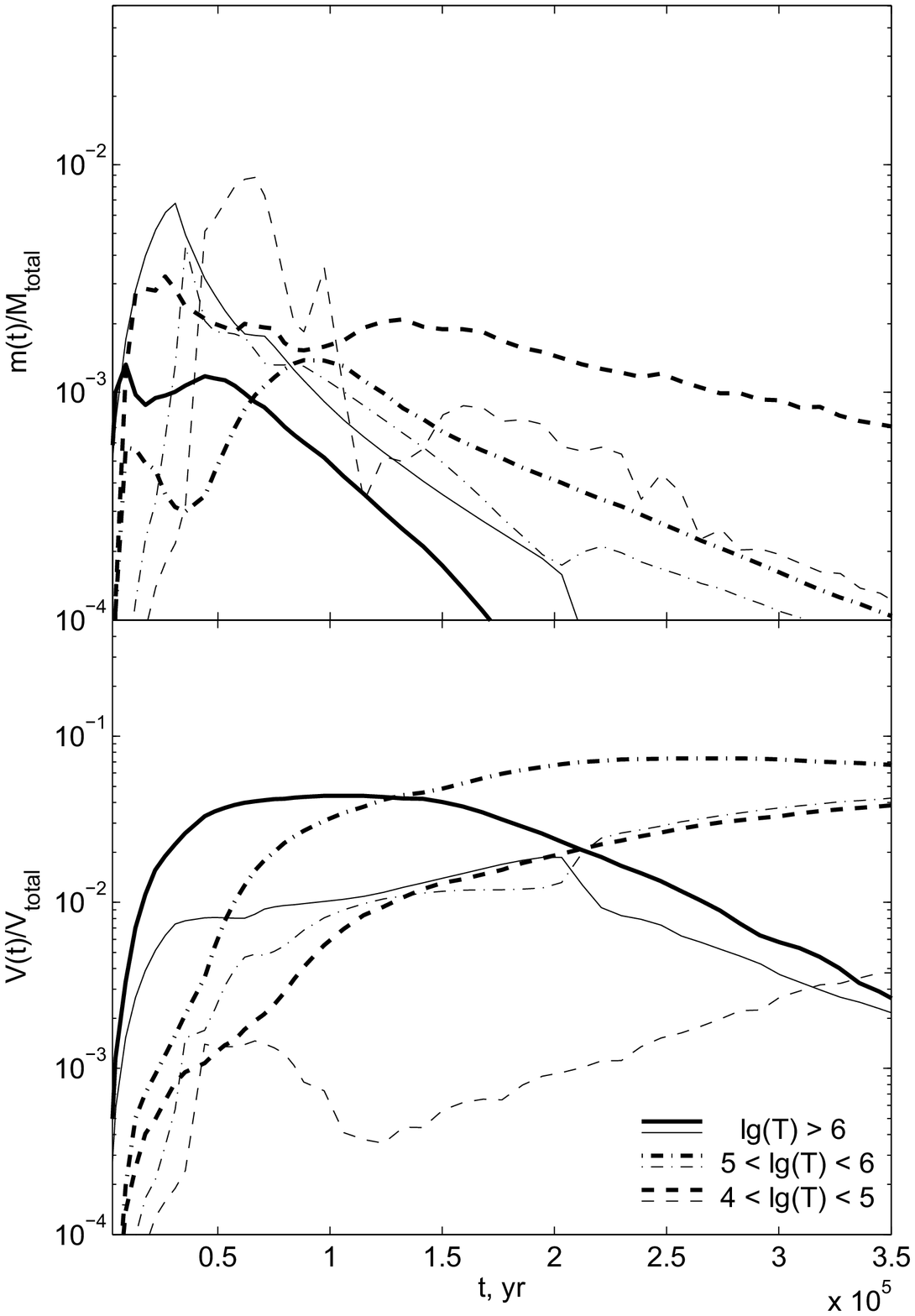}
\caption{
Evolution of the normalized mass (upper row) and volume (lower row) of the gas with temperatures $T>10^6$~K, $T=10^5-10^6$~K, and $T=10^4-10^5$~K (solid, dot-dashed,and dashed curves, respectively) for filling factors $f=0.05$ (left plots) and 0.2 (right plots).
}
\label{evolmass-vol}
\end{figure}
%%%%%%%%%%%%%%%%%%%%%%%%%%%%%%%%%%%%%%%%%%%%%%%%%%%%%%%%%%%%%%%%%%%%%

%%%%%%%%%%%%%%%%%%%%%%%%%%%%%%%%%%%%%%%%%%%%%%%%%%%%%%%%%%%%%%%%%%%%%
\begin{figure}[!ht]
\center
\includegraphics[width=10cm]{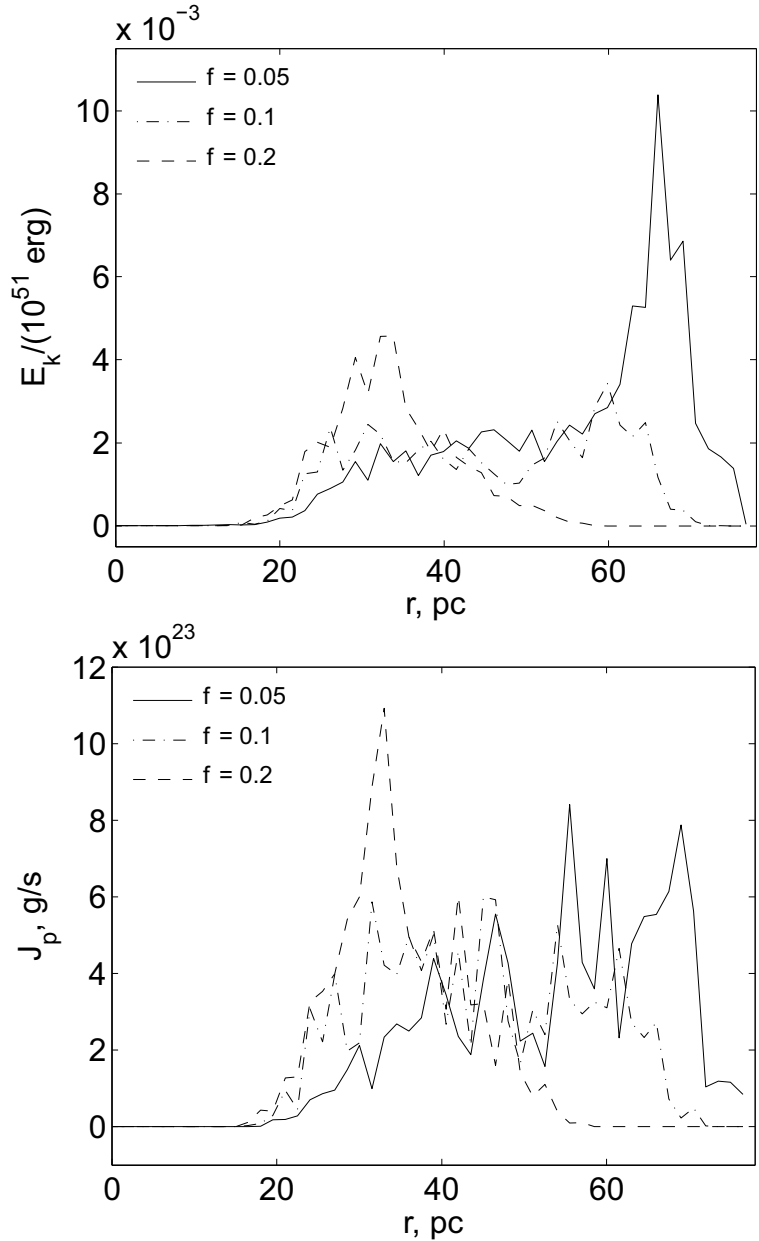}
\caption{
Kinetic energy of an envelope of radius r (upper) and the momentum flux through a surface of radius $r$ (lower) for $f=0.05, 0.1, 0.2$ (solid, dot-dashed, and dashed curves, respectively) for time $t = 3.5 \times 10^5$~yrs after the supernova.
}
\label{evol-kin-mom}
\end{figure}
%%%%%%%%%%%%%%%%%%%%%%%%%%%%%%%%%%%%%%%%%%%%%%%%%%%%%%%%%%%%%%%%%%%%%

The presence of a cloudy phase influences the dynamics of the supernova envelope, especially when the volume filling factor is high ($f\sim
0.1-0.2$): not only the character of the propagation of the supernova shock, but also the dynamics of the gas far behind the shock front -- the gas flow in the entire cavity -- change. The disruption of the clouds adds cool gas to the overall flow, leading to strong fluctuations in the temperature and pressure. Thus, the thermal and dynamical evolution of the hot cavity both change.

We expect that the inhomogeneity of the gas ahead of the shock front will also influence the redistribution of metals ejected by the supernova. A discussion of this process will be presented in a separate study. Here, we note only one result that qualitatively characterizes the efficiency of the mixing of metals. Figure 3 presents the distribution of the gas metallicity for time $t = 3.5\times 10^5$~yrs after supernova that explode in media with filling factors $f = 0.05, 0.1, 0.2$. This clearly shows that, as the filling factor increases, the metals essentially become trapped in the central part of the supernova remnant, where a high metallicity of the gas is preserved. This is obviously a consequence of the formation of the dense layer of disrupted clouds, which contains the bulk of the momentum and kinetic energy of the supernova remnant. Gas with enhanced metallicity penetrates inside this layer in some places, where the metals are rapidly mixed with the low-metal gas due to chaotic flows in the layer. The mass of enriched gas grows with the filling factor.

%----------------------- Section 5 -------------------------------
\section{Observational manifestations}

\noindent

The properties of the dynamical and thermal evolution of a supernova remnant due to the presence of clouds in the ambient medium should certainly be manifest in the characteristics of the associated emission. Consequently, it should be possible to estimate the character of the distribution of gas and metals from the intensity of the energy losses in various spectral ranges (see, e.g., \cite{mckee82,white90}).

Figure 7 shows that the mass of hot gas rapidly falls in all the models considered. Note that the mass of hot gas is higher for an explosion in a uniform medium with density $\chi f \rho$ than for an explosion in a cloudy medium with filling factor $f$, density of
the inter-cloud phase $\rho$, and density of the clouds $\chi\rho$. Consequently, the expected X-ray luminosity is higher for a supernova remnant in a uniform than in a cloudy medium. The increase in the density of the medium facilitates both a decrease in the luminosity and a reduced duration for the bright X-ray phase. The growth of the filling factor in a cloudy medium has the same effect: the maximum total mass of hot gas and X-ray luminosity fall by roughly a factor of two when the filling factor decreases from 0.05 to 0.2,\footnote{The total mass of gas in the computational domain is $5.2 \times 10^4~\msun$ and $1.8\times 10^5~\msun$ for filling factors of 0.05 and 0.2, respectively.} while the duration of the bright X-ray phase is restricted to $t\simlt 1.5\times 10^5$~yrs for $f=0.05$ and $t\simlt 0.75\times 10^5$~yrs for $f=0.2$. For a supernova in a medium with $f=0.05$, the bulk of the gas is in the warm phase ($10^5$K$<T<10^6$K) beginning from $t\sim 10^5$~yrs, while the gas rapidly cools to $T<10^5$~K in a medium with $f=0.2$. Thus, we expect variations in the emission characteristics of a supernova remnant with increasing filling factor (or a growth of density in a uniform medium), for example, in the X-ray range, in lines of highly ionized oxygen, and possibly in recombination lines of hydrogen and helium.

To illustrate this, Fig. 9 presents the evolution of the emission measure of gas with temperature $T>10^5$~K along lines of sight at transverse distances of $r=0$ and 10~pc, parallel to the $z$ axis (to avoid the influence of features due to the distribution of the medium along a single direction, we averaged over ten neighboring lines of sight). In the case of an explosion in a uniform medium, the emission measure at these transverse distances differ only insignificantly. The mass of gas with temperature $T>10^5$~K varies only slightly in the first $(0.3-0.5)\times 10^5$~yrs, while the emission measure remains essentially constant. The emission measure grows with the external density of the gas. The subsequent sharp fall is due to the transition of the envelope to the radiative phase (Figs. 4, 5), which is expressed as a decrease in themass and a sharp saturation of the volumes of the hot and thermal gas (Fig. 7). Further, the emission measure falls monotonically, with the rate of this drop increasing with the density of the ambient medium.

In the case of a supernova in a cloudy medium, the transition to the radiative phase is prolonged in time, due to the lower density of the inter-cloud gas (Figs. 4, 5), extending the period when the emission measure remains nearly constant. The density distribution in the remnant is very non-uniform (Fig. 3), which is expressed in the case of a high filling  factor through the formation of a layer of disrupted cloud fragments (Fig. 8). Since a large number of cloud fragments with temperatures substantially below $10^5$~K lie in the line of sight at $r=10$~pc, after the onset of the radiative phase, the emission measure becomes comparable to (for $f=0.05$) or somewhat lower than (for $f=0.20$) in the case of a uniform medium (Fig. 9). The emission measure falls rapidly at larger transverse distances. Thus, substantial variations in the emission measure of gas with $T>10^5$~K for lines of sight at different transverse distances are expected for a supernova explosion in a cloudy medium.

The deceleration of the supernova envelope should be manifest through variations in the widths of recombination lines of hydrogen and ions of metals. The velocity dispersion of the ionized gas should decrease with increasing age, with this decrease being stronger the more efficient the cooling of the gas. Figure 9 shows the evolution of the velocity dispersion of the ionized gas (assuming that gas with $T>10^4$~K is fully ionized) along lines of sight with transverse distances $r=0$ and 10~pc, parallel to the $z$ axis. After a substantial drop at the beginning of the evolution, the velocity dispersion remains nearly constant up to the onset of the radiative phase. The dispersion grows monotonically in the case of a uniform medium with only a small contribution from a cloudy component, $f\leq 0.05$. In a denser medium, corresponding to a fairly high contribution from clouds, $f>0.05$, the initial increae in the dispersion is replaced by a decrease $t\sim 2\times 10^5$~yrs. This is due to the interaction of the stronger reverse shock and the rapidly cooling supernova envelope. The reverse shock catches up to the forward shock earlier in the case of a uniform medium corresponding to a large contribution from clouds. The dispersion reaches its local maximum at this time, then falls after the interaction of the two shock fronts. The difference between the velocity dispersions for lines of sight corresponding to transverse distances $r=0$ and 10~pc becomes minimum after the onset of the radiative phase. Note that the velocity dispersion for an envelope expanding into a uniform medium is determined by the maximum velocity projected onto the line of sight; i.e., the velocity of the forward or reverse shock.

%%%%%%%%%%%%%%%%%%%%%%%%%%%%%%%%%%%%%%%%%%%%%%%%%%%%%%%%%%%%%%%%%%%%%
\begin{figure}[!ht]
\center
\includegraphics[width=7.5cm]{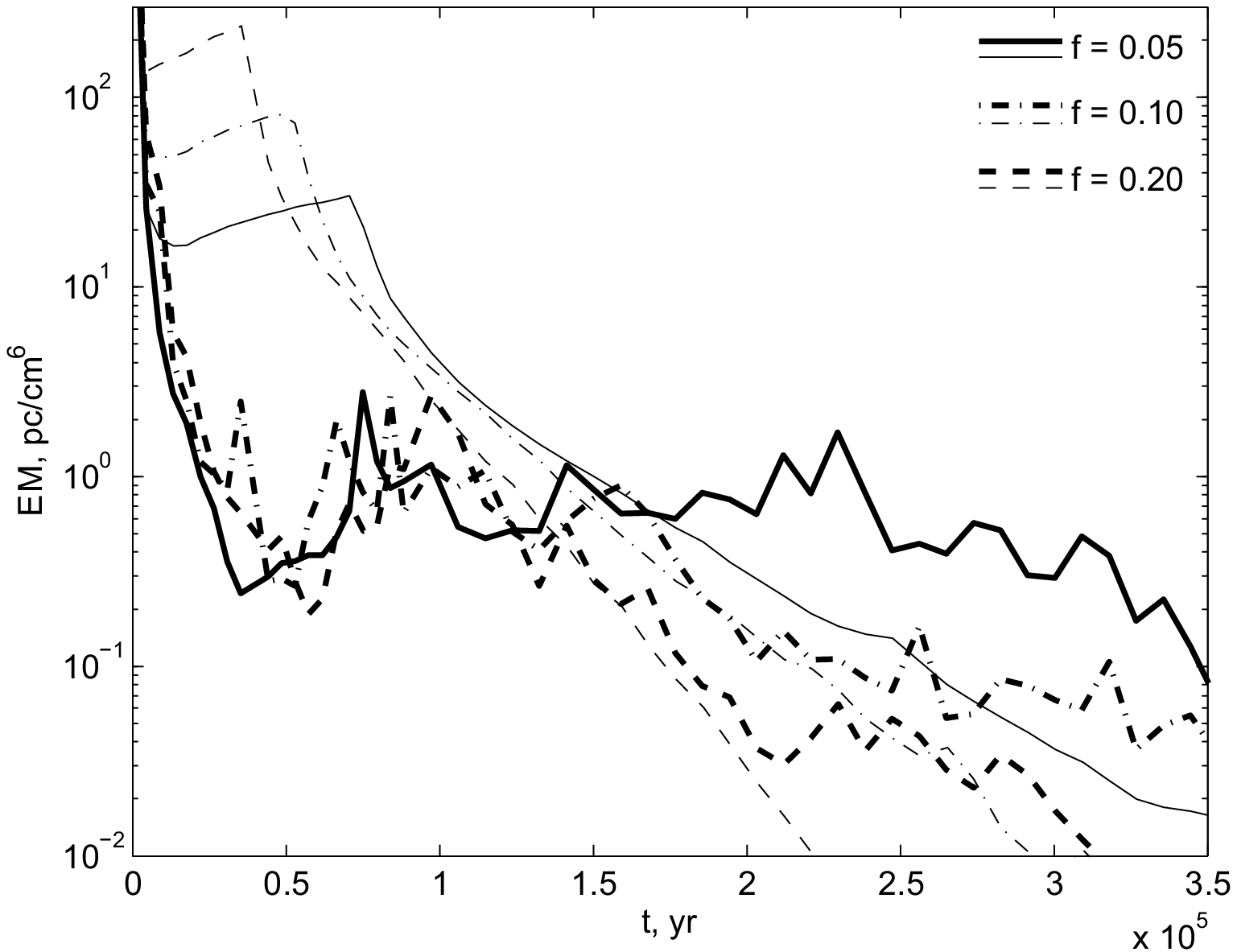}
\includegraphics[width=7.5cm]{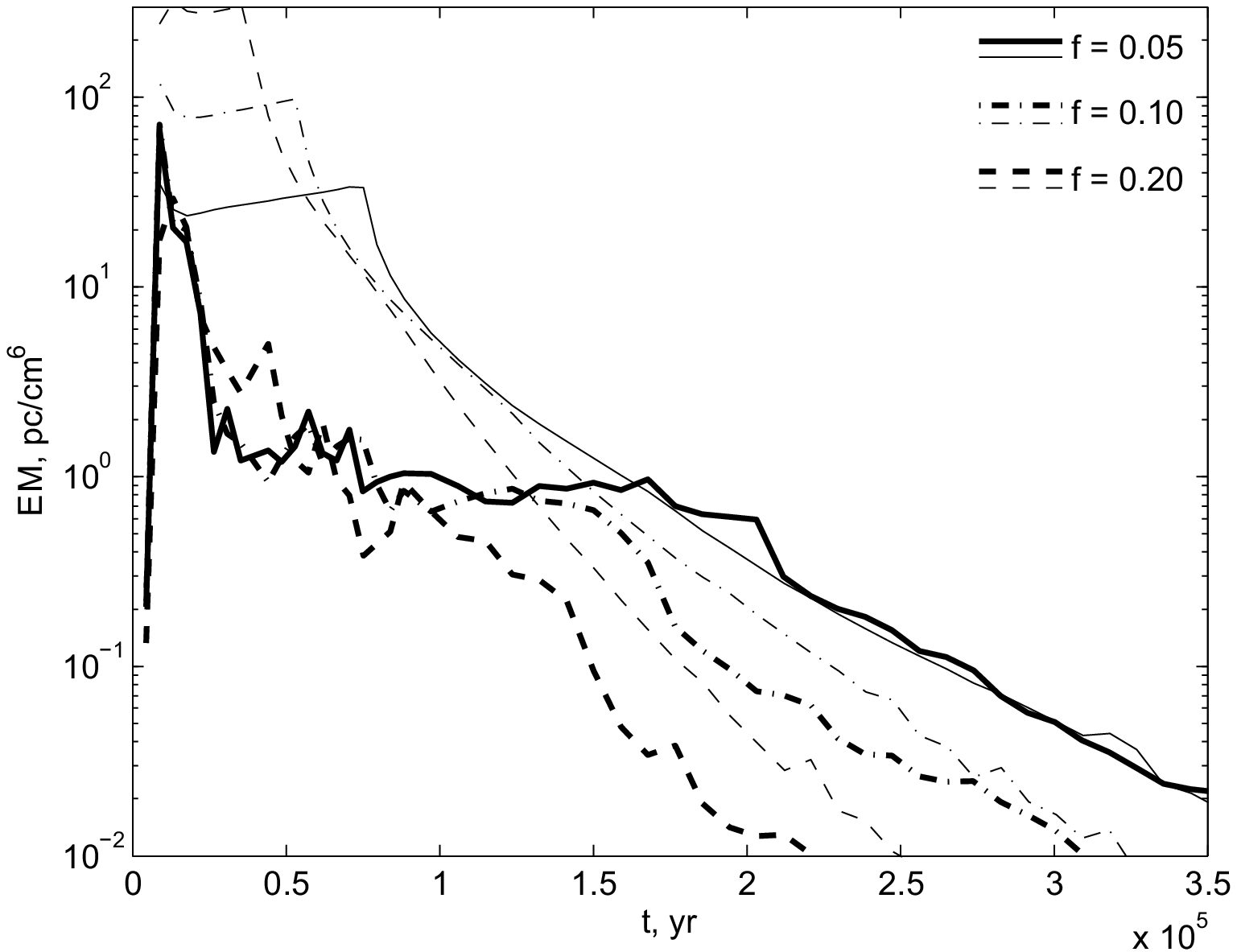}
\includegraphics[width=7.5cm]{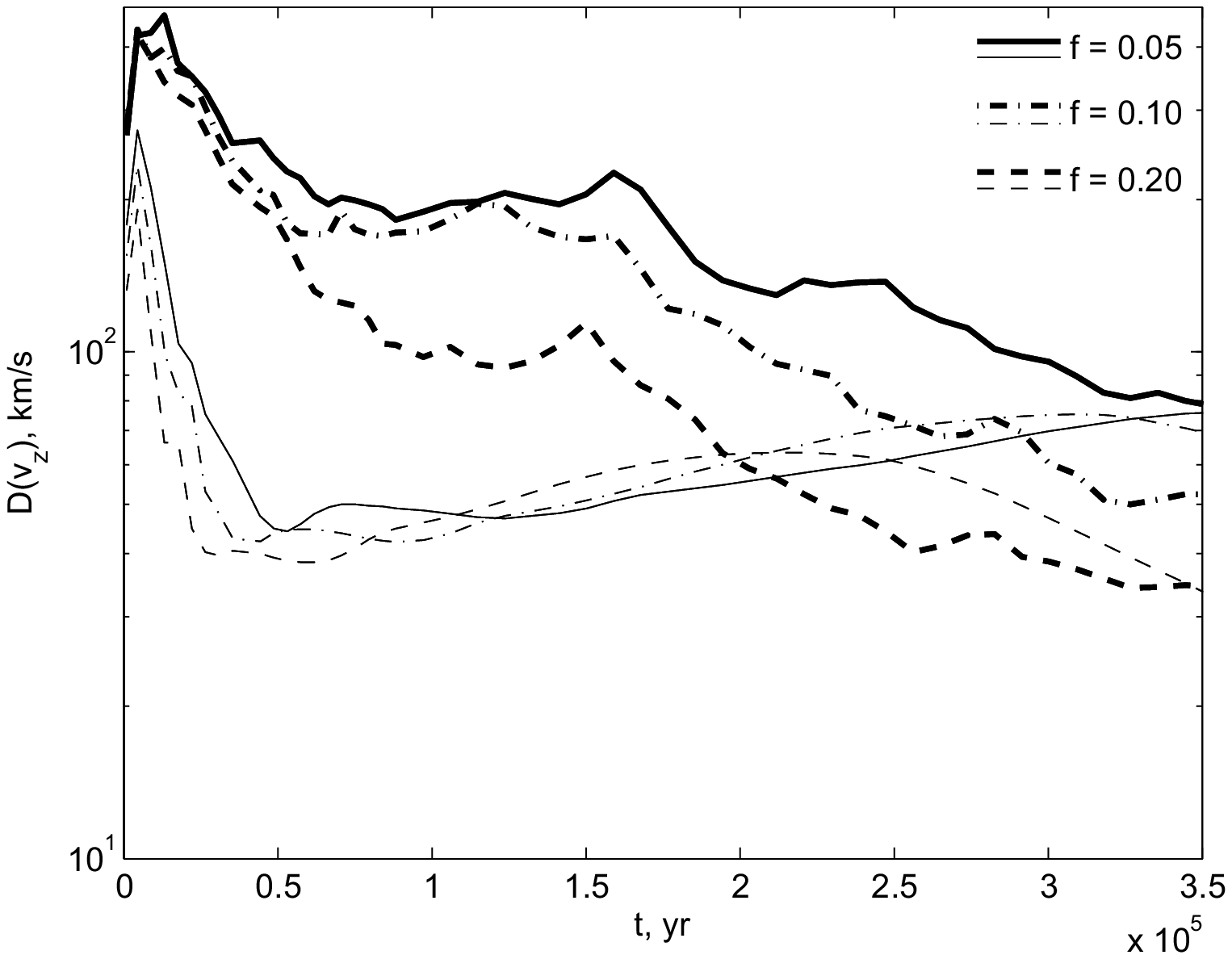}
\includegraphics[width=7.5cm]{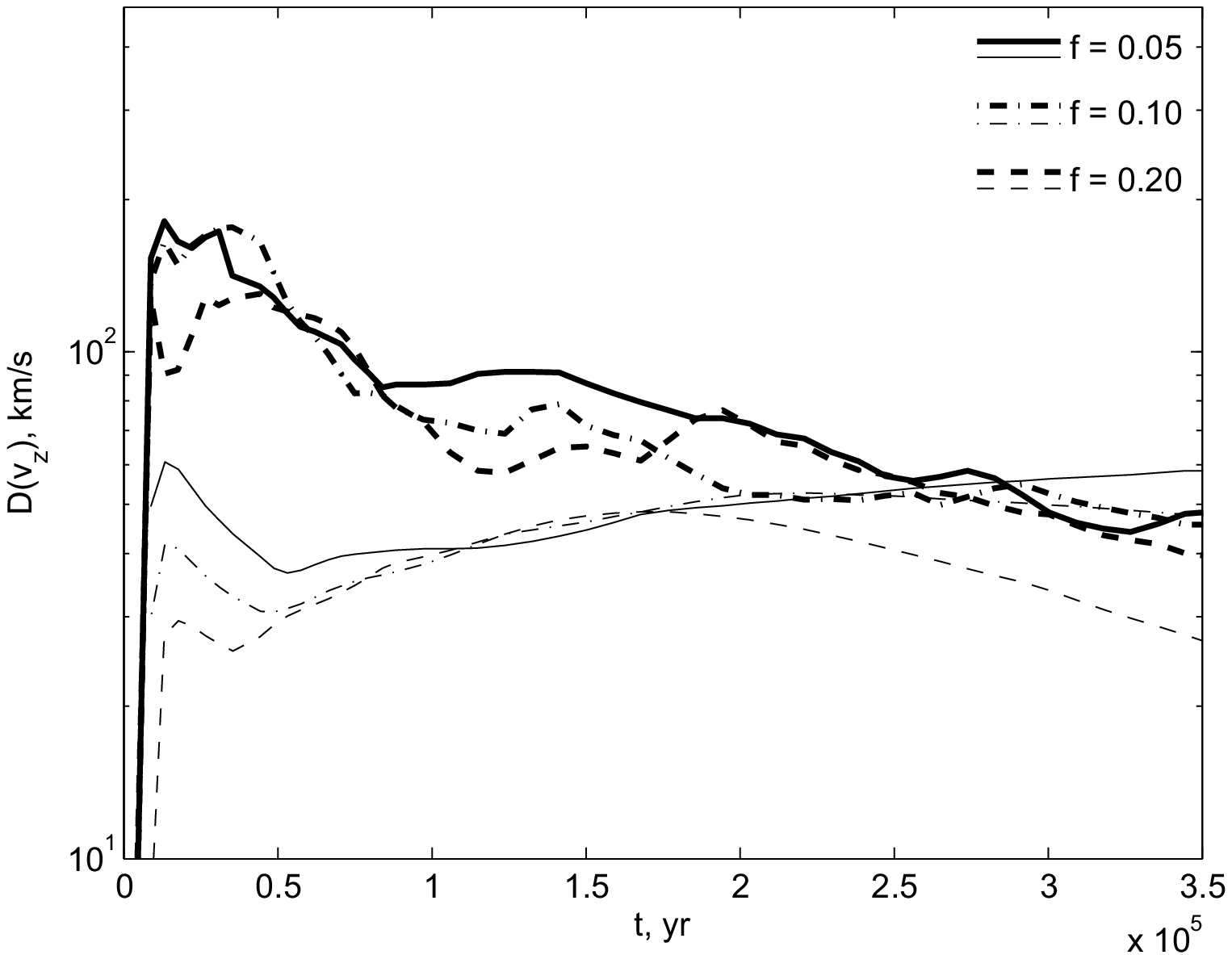}
\caption{
Evolution of the emission measure of the hot and warm ($T>10^5$~K) gas (upper plots) and the velocity dispersion of the ionized gas (lower plots) along the line of sight at transverse distances $r=0$ and 10~pc, parallel to the $z$ axis. The left and right plots correspond to hot and warm–hot gas, respectively.
}
\label{evol-em-vdisp}
\end{figure}
%%%%%%%%%%%%%%%%%%%%%%%%%%%%%%%%%%%%%%%%%%%%%%%%%%%%%%%%%%%%%%%%%%%%%

The velocity dispersion is higher for a supernova in a cloudy medium than in uniform gas. This is due to the low density of the inter-cloud gas, which leads  to a higher expansion velocity of the remnant. Due to the non-uniform distribution of gas both ahead of the front and inside the remnant, the reverse shock is not well centered, so that the velocity dispersion decreases with time, on average. Figure 9 clearly shows that the behavior of the velocity dispersion changes with the transverse distance when $t\simgt 1.5\times 10^5$~yrs: the dispersion depends on the filling factor when $r=0$, but this dependence becomes insignificant in the transition to $r=10$~pc. Note that, for lines of sight that are close to the axis, the maximum velocities are reached in the central region of the cavity of hot gas, which is essentially free of cloud fragments. The flows are close to radial in this region. Since an increase in transverse distance corresponds to a decrease in the velocity projected onto the line of sight in the case of radial flows, this means that the velocity dispersion falls. However, when the filling factor is increased, a layer of disrupted clouds appears; narrow channels filled with hot, rarified gas remain between the dense fragments. The flow of this gas is obviously non-radial, and the velocities in this flow are fairly high. Therefore, the maximum gas velocities at large transverse distances and for high filling factors are reached precisely in the flow of hot gas along channels inside the layer of disrupted clouds. The continual collisions of fragments in this layer lead to a gradual decrease in the velocity dispersion. Thus, appreciable variations of certain quantities, such as the emission measure of the thermal gas and the velocity dispersion, as functions of the volume filling factor and transverse distance can serve as indirect indicators of the character of inhomogeneity in the interstellar medium in which the remnant expands.

%----------------------- Section 5 -------------------------------
\section{Conclusions}

\noindent

We have considered the evolution of a supernova remnant after a supernova explosion in a cloudy medium. We have studied the dependence of the dynamics of the remnant and the physical properties of the gas as functions of the cloud volume filling factor $f$. Our model takes into account mixing of heavy elements ejected into the ambient medium during the supernova explosion. Our studies have shown the following.

\begin{enumerate}
 \item  The presence of clouds leads to a non-simultaneous and early transition of individual parts of the supernova envelope to the
  radiative phase. This transition is characterized by appreciable inhomogeneity of the density and temperature of the gas, both in the
  supernova envelope and inside the hot cavity. Increasing the cloud volume filling factor decreases the time for the transition to the
  radiative phase.
 \item The mean radius of the supernova envelope decreases with growing volume filling factor. The difference in the size of the envelope
  is manifested nearly from the very beginning of its evolution, and this difference reaches factors of about 1.3 and 1.7 by time
  $t = 3.5\times 10^5$~yrs, for filling factors of 0.05 and 0.2, respectively.
 \item When the supernova shock expands into a homogeneous medium with a constant density equal to the mean density $\chi f\rho$ (where
  $\chi$ is the excess of the gas density in clouds over the density $\rho$ in the inter-cloud gas), the size of the supernova envelope 
  is smaller and the transition to the radiative phase earlier than in the case of expansion in a cloudy medium with volume filling 
  factor $f$.
 \item The onset of the radiative stage is accompanied by the development of hydrodynamical instabilities in the supernova envelope. As 
  a consequence, the thermal energy falls off more rapidly, as $E_t\sim t^{-2.3}$, as in the propagation of a supernova remnant through
  either a uniform or a cloudy medium. Beginning from this time, the kinetic energy of the envelope decreases due to the interactions
  between the envelope fragments.
 \item The efficient energy losses of an envelope propagating in a cloudy medium with a high filling factor lead to a decrease in the
  volume occupied by hot gas: the volume of gas with temperatures $T>10^6$~K and $T=10^5-10^6$~K decreases by about a factor of ten
  when $f$ is increased from 0.05 to 0.2, and the mass of gas in these temperature interals falls even more, by a factor of 40–200.
 \item Increasing the cloud volume filling factor leads to an appreciable radial re-distribution of the kinetic energy and momentum 
  between the global supernova shock front and numerous shock fronts that rise during the disruption of the clouds; when $f\simgt 0.1$,
  a gaseous layer with excess kinetic energy and momentum forms behind the global shock front, with the bulk of the kinetic energy and
  momentum in this layer concentrated in fragments of disrupted clouds. This layer with excess kinetic energy and momentum essentially
  traps the hot gas of the cavity in the central region of the supernova remnant when $f\simgt 0.1$. Thanks to this, the initial (high)
  metallicity is essentially preserved in this hot gas.
\end{enumerate}

Thus, the interaction of a supernova envelope with a cloudy interstellar medium appreciably influences the evolution of the entire supernova remnant, more specifically the dynamics and structure of the gas distribution, which are manifest through the observed characteristics of the supernova remnant. In particular, we expect substantial fluctuations of the emission measure of the gas with $T>10^5$~K and the velocity dispersion of the ionized gas, as well as their dependence on the transverse distance, in the case of a supernova explosion in a cloudy medium.

%----------------------- Section K -------------------------------
\section{Acknowledgements}

\noindent

This work was supported by the Russian Foundation for Basic Research (grants 12-02-00365, 12-02-00917, 12-02-92704, 15-42-02682, 15-02-08293), the 'Dinasty' Foundation, and the Ministry of Education and Science of the Russian Federation (project 213.01-11/2014-5).

%----------------------- Section L -------------------------------

%----------------------- Figures -------------------------------

\end{document}